\documentclass[aps,prb,groupedaddress,floats,preprint]{revtex4-2}

\usepackage[english]{babel}

\usepackage{amsfonts,graphicx,soul,natbib,mathrsfs}

\usepackage[colorlinks=true]{hyperref}
\usepackage[dvipsnames]{xcolor}
\usepackage[normalem]{ulem}
\usepackage{epstopdf}
\usepackage{textcomp}
\usepackage{xcolor}
\usepackage{soul}
\soulregister\cite7
\soulregister\ref7
\soulregister\eqref7
\usepackage{physics}
\usepackage{graphics}
\usepackage{comment}
\usepackage{amssymb}
\usepackage{tensor}
\usepackage{graphicx,subcaption}
\usepackage[dvipsnames]{xcolor}

\newcommand{\expv}[1]{\left\langle #1 \right\rangle}

\usepackage{dcolumn}
\usepackage{bm}

\begin{document}

\title{Nonstationary Laguerre-Gaussian states in magnetic field}

\author{G.K. Sizykh}
\email{georgii.sizykh@metalab.ifmo.ru}
\affiliation{School of Physics and Engineering,
ITMO University, Lomonosova 9, St. Petersburg, 197101, Russia}

\author{A.\,D.~Chaikovskaia}
\affiliation{School of Physics and Engineering, ITMO University, Lomonosova 9, St. Petersburg, 197101, Russia}

\author{D.V. Grosman}
\affiliation{School of Physics and Engineering, ITMO University, Lomonosova 9, St. Petersburg, 197101, Russia}

\author{I.I. Pavlov}
\affiliation{School of Physics and Engineering, ITMO University, Lomonosova 9, St. Petersburg, 197101, Russia}

\author{D.V. Karlovets}
\email{dmitry.karlovets@metalab.ifmo.ru}
\affiliation{School of Physics and Engineering, ITMO University, Lomonosova 9, St. Petersburg, 197101, Russia}

\begin{abstract}
The Landau states of electrons with orbital angular momentum in magnetic fields are important in the quantum theories of metals and of synchrotron radiation at storage rings, in relativistic astrophysics of neutron stars, and in many other areas. In realistic scenarios, electrons are often born inside the field or injected from a field-free region, requiring nonstationary quantum states to account for boundary or initial conditions. This study presents nonstationary Laguerre-Gaussian (NSLG) states in a longitudinal magnetic field, characterizing vortex electrons after their transfer from vacuum to the field. Comparisons with Landau states and calculations of observables such as mean energy and r.m.s. radius show that the r.m.s. radius of the electron packet in the NSLG state oscillates in time around a significantly larger value than that of the Landau state. This quantum effect of oscillations is due to boundary conditions and can potentially be observed in various problems, particularly when using magnetic lenses of electron microscopes and linear accelerators. Analogies are drawn between a quantum wave packet and a classical beam of many particles in phase space, including the calculation of mean emittance of the NSLG state as a measure of their quantum nature.
\end{abstract}

\maketitle


\section{Introduction}
During the last two decades, freely propagating electrons with orbital angular momentum (OAM), dubbed twisted or vortex electrons, have successfully transitioned from a theoretical concept \cite{Bliokh2007, Bliokh2011, Bliokh2012, Gallatin2012, Karlovets2012, Greenshields2012, Ivanov2016, Karlovets2018, Maiorova2018, Ducharme2021, Karlovets2021Nonlinear, Shaohu2021} to experimental realizations \cite{Verbeeck2010, McMorran2011, Guzzinati2013, Petersen2013, Schattschneider2014, Schachinger2015} and possible practical implementations \cite{Verbeeck2010, Idrobo2011, Mohammadi2012, Grillo2017}. Nevertheless, this is still a relatively new area in electron microscopy and particle physics \cite{Bliokh2017}. In particular, generation and lensing of twisted electrons should be thoroughly investigated so they could become a reliable and useful tool in atomic and particle physics, studies of magnetic properties of materials \cite{Idrobo2011, Mohammadi2012}, and other associated fields.

The nonstationary vortex electron states are close relatives of the widely known stationary Landau states in a magnetic field, because both can carry a quantized orbital angular momentum with respect to the mean propagation direction. However, these two types of solutions do not continuously transform into each other when dealing with realistic problems where boundaries appear, for instance, the electron entering the field from a field-free region or being generated inside the field at some moment of time. While the common approach is to assume that the electron evolves to the Landau state right after entering the field \cite{Karlovets2021Vortex, Greenshields2014, Greenshields2015}, it seems highly unlikely that this transition lasts an infinitesimal period of time. Therefore, this common approach appears to have limited applicability. 

The Landau states are also widely used to describe the dynamics and radiation of quantum particles in synchrotrons \cite{STeng, BagrovBlackBook}. These states rely on the assumption of an infinite particle lifetime within an accelerating ring. However, considering more general nonstationary states allows for arbitrary time intervals, thereby providing insights into the quantum theory of radiation from a short sector of a synchrotron and offering a more accurate description of particle dynamics in these setups. 

Additionally, one can set the problem of an electron in a constant and homogeneous magnetic field using one of the two distinct gauges for the vector potential ${\bm A}$, both leading to the same field ${\bm H}=\{0,0,H\}$ \cite{Landau}, but to different sets of solutions: namely, the Hermite-Gaussian states or the Laguerre-Gaussian ones. Clearly, these are two distinct physical states with different quantum numbers, and it is the boundary (or initial) conditions that determine the choice of the gauge and of the electron quantum state.

\begin{figure}[t]
\center{\includegraphics[width=0.8\linewidth]{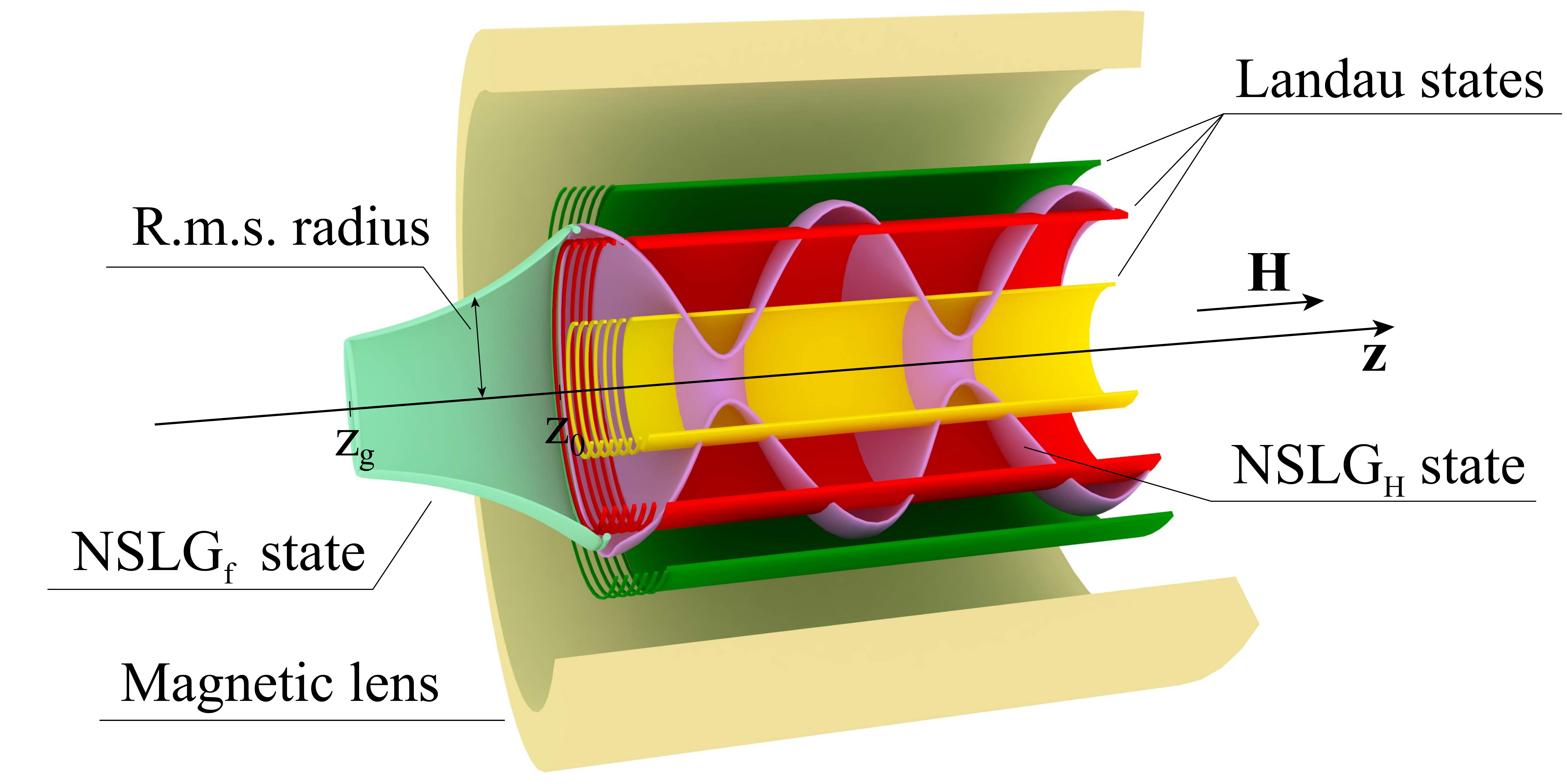}}
\caption{Transfer of a free Laguerre-Gaussian electron through a magnetic lens. $z_{\text{g}}$ and $z_0$ are the positions of the electron source and the boundary, respectively.} 
\label{fig:Lens3D}
\end{figure}

Here we argue that it is generally {\it the nonstationary Laguerre-Gaussian} (NSLG) states rather than the Landau ones that correctly describe the process of transition of an electron from vacuum to the longitudinal magnetic field with the appropriate boundary conditions. Introducing a boundary makes the root-mean-square (r.m.s) radius of the electron oscillate around a value significantly larger than that predicted by the stationary Landau states. Not only is this solution more realistic than the Landau state because it takes the boundary or initial conditions into account, but control over the electron transfer through magnetic lenses is crucial for the further use of these electrons in electron microscopes and linear accelerators. There have already been attempts to investigate the propagation of electrons carrying OAM in magnetic fields \cite{Bliokh2012, Gallatin2012, Silenko2021, Karlovets2021Vortex, Melkani2021, Baturin2022}; however, in these works little attention is paid to the boundary transition that is essential for the electron dynamics. In this paper, we demonstrate that the transfer of a vortex electron through a boundary between vacuum and a solenoid (in a setup similar to that of Fig. \ref{fig:Lens3D}) can quantitatively be described using the NSLG states.

The aim of this paper is to elaborate on the nonstationary dynamics of electrons in a magnetic field and to investigate the NSLG states in detail. In Sec. \ref{secNSLG}, we introduce these states and provide their comprehensive description both in free space and in a magnetic field. We focus on the electron transverse dynamics, as the longitudinal one is not affected by the magnetic field. The transverse dynamics is supposed to be nonrelativistic and the  restrictions imposed are discussed in Sec. \ref{secRelativism}. 
In Sec. \ref{secNSLGandFree}, we show that in the limit of $H \rightarrow 0$ the NSLG states inside the solenoid turn into free-space Laguerre-Gaussian wave packets. Further, we consider a mismatch between a free NSLG electron propagation axis and the magnetic field direction.  In Sec. \ref{secNSLGandLandau}, the NSLG and the Landau states are compared, particularly, their sizes. Then we decompose the former into the superposition of the latter. Finally, in Sec. \ref{secEmittance}, analogies are drawn between a classical particle beam and a quantum wave packet. We introduce a quantum r.m.s. emittance and apply it to the NSLG states.

Electron spin has no qualitative impact on our results and is neglected. Throughout the paper, natural system of units $\hbar = c = 1$ is used. The electron charge is $e = - e_0$, where $e_0 > 0$ is the elementary charge. Alongside with the electron mass, we use the Compton wavelength $\lambda_{\text{C}} = m^{-1}$.

\section{NSLG states}
\label{secNSLG}

\subsection{Longitudinal and transverse dynamics}

In nonrelativistic quantum mechanics, electron dynamics is described by the Schr\"{o}dinger equation
\begin{equation}
\label{Schr}
    i \frac{\partial \Psi (\bm{r},t)}{\partial t} = \hat{\mathcal H} \Psi(\bm{r},t).
\end{equation}
Both in vacuum and inside a magnetic lens, we can single out the motion along the field and factorize the solution of Eq. \eqref{Schr} as $\Psi(\bm{r},t) = \Psi_{\perp}(\rho, \varphi, t) \Psi_{\parallel}(z, t)$. 

The longitudinal wave function is assumed to be a wave-packet solution to the one-dimensional Schr\"{o}dinger equation
\begin{equation}
\label{SchrPar}
    i \frac{\partial \Psi_{\parallel}}{\partial t} = \frac{\hat{p}_z^2}{2m} \Psi_{\parallel}
\end{equation}
with a nonzero average $z$-projection of the velocity operator $-i\lambda_{\text{C}}\expv{\partial_z} = v$. Generally, it can be presented as a superposition of plane waves with different momenta:
\begin{equation}
\label{lo}
\Psi_{\parallel}(z,t) = \int\limits_{-\infty}^{\infty} g(p_z)\exp(i p_z z - i \frac{p_z^2}{2m} t)\frac{dp_z}{2\pi}.
\end{equation}
Its explicit form does not affect the transverse dynamics. From here on, we only discuss the transverse dynamics of twisted electrons and omit the ``$\perp$'' sign to simplify the notation.

\subsection{General nonstationary Laguerre-Gaussian states}

In the present work, we are interested in the transverse dynamics of an electron after it crosses the boundary between vacuum and a magnetic field area. In both regions, the electron can be described by the following wave function:
\begin{equation}
\label{NSLG}
    \Psi_{n l}(\bm{\rho},t) = N_{n l} \frac{\rho^{|l|}}{\sigma^{|l|+1}(t)} L_n^{|l|} \left(\frac{\rho^2}{\sigma^2(t)}\right) \exp \left[il\varphi - i\Phi_{\text{G}}(t) -  \frac{\rho^2}{2\sigma^2(t)}\left(1-i\frac{\sigma^2(t)}{\lambda_{\text{C}} R(t)}\right)\right],
\end{equation}
which we call a \textit{nonstationary Laguerre-Gaussian} state. Here, $L_n^{|l|}$ are generalized Laguerre polynomials, $n=0,1,2,...$ is the radial quantum number, and $l = 0,\pm 1,\pm 2,...$ is the OAM, which is conserved in axially symmetric fields even with weak inhomogeneities \cite{Karlovets2021Vortex}. The difference between $\operatorname{NSLG}$ states in free space ($\operatorname{NSLG}_{\text{f}}$) and in the magnetic field ($\operatorname{NSLG}_{\text{H}}$) is determined by optical functions: dispersion $\sigma(t)$, radius of curvature $R(t)$, and Gouy phase $\Phi_{\text{G}}(t)$. The normalization constant in Eq. \eqref{NSLG} is defined by the standard condition of a single particle in the volume:
\begin{equation}
\label{Norm}
    N_{n l} = \sqrt{\frac{1}{\pi} \frac{n!}{(n + \abs{l})!}}.
\end{equation}
The NSLG states were briefly introduced in our recent work \cite{Sizykh2023} as means to account for the boundary crossing that provide consistent description of the electron state in regions with and without magnetic field. Here we dwell deeper into the dynamics of these states and discuss their properties from different angles.

The state with the transverse part \eqref{NSLG} corresponds to an electron moving rectilinearly along the $z$-axis, which means that \begin{equation}
    \expv{\bm{\rho}} = 0, \quad \expv{\hat{\bm{v}}} = 0,
\end{equation}
where $\hat{\bm{v}} = - i \nabla_{\perp} / m - e \bm{A} / m$. The 
r.m.s. radius  of the NSLG state is proportional to the dispersion:
\begin{equation}
\label{r.m.s.Rad}
    \rho(t) \equiv \sqrt{\expv{\rho^2} - \expv{\bm{\rho}}^2}  = \sigma(t) \sqrt{2n+|l|+1}.
\end{equation}

We can directly check that states \eqref{NSLG} form an orthonormal set:
\begin{equation}
    \label{Orthonormality}
    \int \Psi^*_{n' l'}(\bm{\rho},t) \Psi_{n l}(\bm{\rho},t) d^2 \rho = \delta_{n n'} \delta_{l l'}.
\end{equation}
The set is also complete (see the proof in the Appendix~\ref{sec:Completeness}).

\subsection{Nonstationary Laguerre-Gaussian states in free space}

In this section, we derive the optical functions of the $\operatorname{NSLG}_{\text{f}}$ states, which will later determine the initial conditions for the states in the field.

In free space, the transverse Hamiltonian is
\begin{equation}
\label{HamFree}
    \hat{\mathcal H}_{\text{f}} = \frac{\hat{\bm{p}}_{\perp}^2}{2m},
\end{equation}
where the index ``f'' stands for ``free''. To derive the optical functions and then the $\operatorname{NSLG}_{\text{f}}$ state, the wave function \eqref{NSLG} can be substituted into the Schr\"{o}dinger equation \eqref{Schr} with the Hamiltonian \eqref{HamFree}. This leads to the system of equations
\begin{equation}
\label{OptEqFree}
    \begin{aligned}
        &\frac{1}{R(t)} = \frac{\sigma'(t)}{\sigma(t)},\\
        &\frac{1}{\lambda_{\text{C}}^2 R^2(t)} + \frac{1}{\lambda_{\text{C}}^2} \left[\frac{1}{R(t)}\right]' = \frac{1}{\sigma^4(t)},\\
        &\frac{1}{\lambda_{\text{C}}}\Phi_{\text{G}}'(t) = \frac{2n + |l| + 1}{\sigma^2(t)},
    \end{aligned}
\end{equation}
where the primes stand for time derivatives. Instead of $R(t)$, we prefer using the dispersion divergence rate $\sigma'(t) = \sigma(t) / R(t)$ alongside with $\sigma(t)$ and $\Phi(t)$ to characterize the NSLG states.

To find the unique solution of the system \eqref{OptEqFree}, the initial conditions should be specified. In real experiment, twisted electrons are generated at the beam waist:
\begin{equation}
\label{InConFree}
    \sigma_{\text{f}}(t_{\text{g}}) = \sigma_{\text{w}}, \quad \sigma'_{\text{f}}(t_{\text{g}}) = 0, \quad \Phi_{\text{f}}(t_{\text{g}}) = 0,
\end{equation}
where $t_{\text{g}}$ is the time when the twisted electron is generated and $\sigma_{\text{w}}$ is the dispersion at the waist. We set $\Phi_{\text{f}}(t_{\text{g}}) = 0$, because a constant phase factor does not change the state.

The optical functions $\sigma_{\text{f}}(t)$ and $\Phi_{\text{f}}(t)$ satisfying the system \eqref{OptEqFree} with the initial conditions \eqref{InConFree} are
\begin{equation}
\label{OptSolFree}
    \sigma_{\text{f}}(t) = \sigma_{\text{w}} \sqrt{1 + \frac{(t - t_{\text{g}})^2}{\tau_{\text{d}}^2}}, \hspace{30pt} \Phi_{\text{f}}(t) = (2n + \abs{l} + 1) \arctan{\left( \frac{t - t_{\text{g}}}{\tau_{\text{d}}} \right)}.
\end{equation}
Here, $\tau_{\text{d}} = \sigma_{\text{w}}^2 / \lambda_{\text{C}}$ is the diffraction time. The NSLG states \eqref{NSLG} with $\sigma(t)$ and $\Phi_{\text{G}}(t)$ given by Eqs. \eqref{OptSolFree} and $R(t) = \sigma_{\text{f}}(t) / \sigma'_{\text{f}}(t)$ are the nonstationary counterparts \cite{Karlovets2019Intrinsic,Karlovets2019Dynamical,Karlovets2021Vortex} of the well-known paraxial \textit{free Laguerre-Gaussian} wave packets \cite{McMorran2011, Bliokh2012, Guzzinati2013, Schachinger2015, Bliokh2017}.

According to Eqs. \eqref{OptSolFree} and \eqref{r.m.s.Rad}, the r.m.s. radius of the $\operatorname{NSLG}_{\text{f}}$ state is
\begin{equation}
\label{r.m.s.RadFree}
    \rho_{\text{f}}(t) = \rho_{\text{w}} \sqrt{1 + \frac{(t - t_{\text{g}})^2}{\tau_{\text{d}}^2}}
\end{equation}
where $\rho_{\text{w}} = \sigma_{\text{w}} \sqrt{2n + \abs{l} + 1}$. This expression illustrates quadratic divergence of the r.m.s. radius near the beam waist and linear growth far from it.

Since the $\operatorname{NSLG}$ states do not generally possess definite energy, we consider its expectation value. For the $\operatorname{NSLG}_{\text{f}}$ state given by Eqs. \eqref{NSLG} and \eqref{OptSolFree}, taking into account $R(t) = \sigma(t) / \sigma'(t)$, \begin{equation}
\label{FLGEn}
    \expv{E}_{\text{f}} =  \frac{2n + \abs{l} + 1}{2 \lambda_{\text{C}}} \left( \frac{\lambda_{\text{C}}^2 }{\sigma_{\text{f}}^2(t)} + {\sigma'}_{\text{f}}^{2}(t) \right).
\end{equation}
The first term in Eq. \eqref{FLGEn} stems from the size effect and decreases with the volume occupied by the wave packet. The second term has a kinetic nature and is responsible for the radial divergence of the state. The free Hamiltonian \eqref{HamFree} does not depend on time, which means that the average energy is constant. Indeed, by substituting the dispersion \eqref{OptSolFree} and its derivative into Eq. \eqref{FLGEn}, we obtain
\begin{equation}
\label{FLGEnEx}
    \expv{E}_{\text{f}} = \frac{2n + |l| + 1}{2 \tau_{\text{d}}}.
\end{equation}

\begin{figure}[b]
\center{\includegraphics[width=0.8\linewidth]{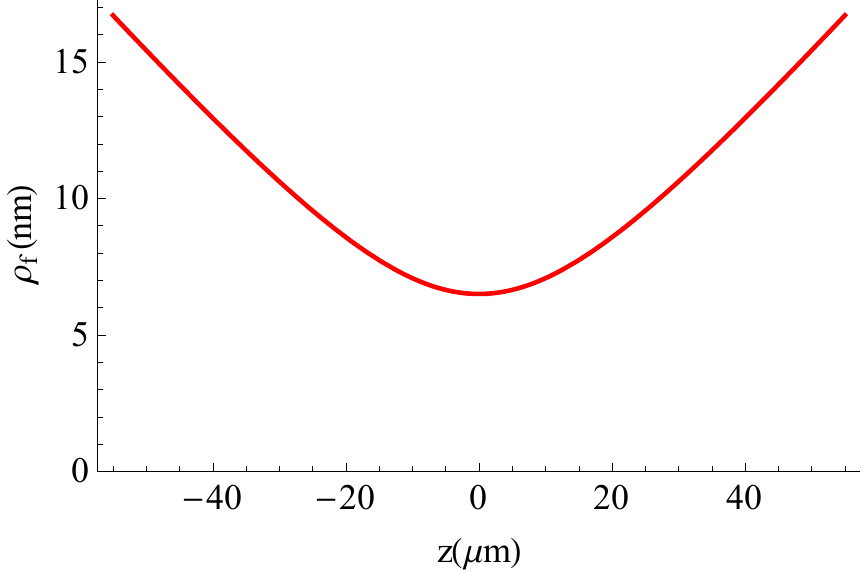}}
\caption{$\operatorname{NSLG}_{\text{f}}$ packet r.m.s. radius based on the parameters from the experiment \cite{Guzzinati2013}. Electron energy $E_{\parallel} = 300 $ KeV (corresponding velocity $v \approx 0.78 c$), $n = 0,~l = 3, $ and $\sigma_{\text{g}} = 3.25$ nm (corresponding $\tau_{\text{d}} = 9 \times 10^{-5}$ ns).}
\label{fig:FreeDisp}
\end{figure}

We illustrate the dynamics of the $\operatorname{NSLG}_{\text{f}}$ wave packet obtained in the experiment of Guzzinati et al. \cite{Guzzinati2013} (see Figs. 3, 4 there) in Fig. \ref{fig:FreeDisp}. The electron has the following parameters: electron energy $E_{\parallel} = 300$ KeV (and the corresponding velocity $v \approx 0.78 c$), $n = 0$, $l = 3$ (in \cite{Guzzinati2013} $l$ is designated as $m$), beam waist dispersion $\sigma_{\text{w}} = 3.25$ nm (corresponding r.m.s. radius of the waist $\rho_{\text{w}} = \sqrt{2n + \abs{l} + 1} = 6.5$ nm), and diffraction time $\tau_{\text{d}} = 9 \times 10^{-5}$ ns.

Note that we plot the beam radius, while in the work \cite{Guzzinati2013} (see Fig. 4(a) there), the beam diameter is depicted. Guzzinati et al. observed several rings as they blocked half of the initial $\operatorname{NSLG}_{\text{f}}$ beam and obtained a superposition of the $\operatorname{NSLG}_{\text{f}}$ states. However, in this case, the original $\operatorname{NSLG}_{\text{f}}$ state makes the dominant contribution, which allows us to reproduce their results.

\subsection{Landau states}

Let us now turn to a twisted electron state inside a solenoid. We describe the solenoid as a semi-infinite stationary and homogeneous magnetic field $\bm{H} = H \theta(z - z_0) \bm{e}_z,\, \bm{e}_z = (0,0,1)$. The step function $\theta(z)$ reflects the hard-edge boundary located at $z_0$. We assume the longitudinal part of the wave function to be narrow enough, so that the field can be considered to be suddenly switched on at the time $t_0$.

Before moving to the $\operatorname{NSLG}_{\text{H}}$ states, we would like to briefly remind the reader of the Landau ones. They are stationary solutions to the Schr\"{o}dinger equation \eqref{Schr} with the transverse Hamiltonian
\begin{equation}
\label{HamField}
    \hat{\mathcal H} = \frac{(\hat{\bm{p}}_{\perp} - e \bm{A})^2}{2m}.
\end{equation}
Recall the aforementioned gauge issue: in the original work of Landau, the vector potential is chosen as \cite{Landau1930}
\begin{equation}
\label{LandauGauge}
    \bm{A} = - H y \bm{e}_x.
\end{equation}
The Landau states that are the solutions of the Schr\"{o}dinger equation \eqref{Schr} with the Hamiltonian \eqref{HamField} defined by the vector potential in the Landau gauge \eqref{LandauGauge} are given by Hermite-Gaussian functions
\begin{equation}
\label{LandauHermite}
    \Psi (x, y, z, t) \propto H_s \left( \frac{y - \tilde{\sigma}_{\text{L}}^2 p_x}{\tilde{\sigma}_{\text{L}}} \right) \exp(- \frac{(y - \tilde{\sigma}_{\text{L}}^2 p_x)^2}{2 \tilde{\sigma}_{\text{L}}^2}) \exp( i p_x x + i p_z z - i \frac{\omega}{2} (2s + 1)),
\end{equation}
where $\tilde{\sigma}_{\text{L}} = \sqrt{1 / \abs{e H}}$, $\omega = \abs{e H} / m$ is the cyclotron frequency, and $s = 0, 1, 2, ...$ is the principal quantum number.

Alternatively, one can choose the symmetric gauge for the vector potential:
\begin{equation}
\label{SymGauge}
    \bm{A} = \frac{H}{2} \rho \bm{e}_{\varphi}.
\end{equation}
where $\bm{e}_{\varphi} = \bm{e}_y \cos{\varphi} - \bm{e}_x \sin{\varphi}$ is the azimuthal unit vector. Such a choice preserves the axial symmetry of the problem, and the corresponding solutions of the Schr\"{o}dinger equation have definite values of the OAM (see, e.g., \cite{Ciftja2020}):
\begin{equation}
\label{Landau}
    \Psi^{(\text{L})}_{n l} (\rho, \varphi,t) = N_{n l} \left( \frac{\rho^{|l|}}{\sigma_{\text{L}}^{|l| + 1}} \right) L_n^{|l|} \left[ \frac{\rho^2}{\sigma_{\text{L}}^2} \right] \exp \left[ -\frac{\rho^2}{2 \sigma_{\text{L}}^2} + i l \varphi - i E_{\text{L}} t \right],
\end{equation}
where $\sigma_{\text{L}} = \sqrt{2 / \abs{e H}}$ is the r.m.s. radius of the Landau state with $n = l = 0$. The normalization constant $N_{n l}$ in Eq. \eqref{Landau} is given by Eq. \eqref{Norm}. In what follows, by Landau states, we mean the wave function \eqref{Landau} and not \eqref{LandauHermite}, which can be viewed as yet another initial condition.

The energy $E_{\text{L}}$ of the Landau states is
\begin{equation}
\label{LandauEn}
    E_{\text{L}} = \frac{\omega}{2} (2n + \abs{l} + l + 1) = \frac{\omega}{2}\left(2n +|l| +1 \right) + l \mu_\text{B} H,
\end{equation}
where $\mu_\text{B} = \abs{e} / (2 m)$ is the Bohr magneton. The last term in Eq. \eqref{LandauEn} is the energy of the magnetic moment $-l \mu_\text{B}$ in the field $H$. Note that the electron energy in a Landau state is infinitely degenerate for $l \leq 0$  due to the exact compensation of kinetic and magnetic ``orbital motions''. However, for $l > 0$, the two terms add up and double the contribution to the energy.

The r.m.s. radius of the Landau states \eqref{Landau} is constant and equal to\begin{equation}
\label{r.m.s.Landau}
    \rho_{\text{L}} = \sigma_{\text{L}} \sqrt{2n + \abs{l} + 1}.
\end{equation}
Note that in a given magnetic field, there is only a countable set of possible r.m.s. radii of an electron described by the Landau states. In reality, an electron enters the field from free space or is generated in the field with an arbitrary size that must evolve continuously. If this size does not fall within the countable set of possible r.m.s. radii, the free electron cannot find a suitable Landau state to transform into. Moreover, even if the r.m.s. radius of the electron equals that of the Landau state, the divergence rate must also vanish. Thus, taking into account the initial conditions, we are generally led to a nonstationary electron state in the field, which is properly described by the $\operatorname{NSLG}_{\text{H}}$ state.

\subsection{Nonstationary Laguerre-Gaussian states in the field}
\label{E}

Similarly to the $\operatorname{NSLG}_{\text{f}}$, one can derive the $\operatorname{NSLG}_{\text{H}}$ states in the magnetic field. Substituting the state \eqref{NSLG} into the Schr\"{o}dinger equation \eqref{Schr} with the Hamiltonian \eqref{HamField} we obtain
\begin{equation}
\label{eq:Opt}
    \begin{aligned}
        &\frac{1}{R(t)} = \frac{\sigma'(t)}{\sigma(t)},\\
        &\frac{1}{\lambda_{\text{C}}^2 R^2(t)} + \frac{1}{\lambda_{\text{C}}^2} \left[\frac{1}{R(t)}\right]' = \frac{1}{\sigma^4(t)} - \frac{1}{\sigma_{\text{L}}^4},\\
        & 
    \frac{1}{\lambda_{\text{C}}}\Phi_{\text{G}}'(t) = \frac{l}{\sigma_{\text{L}}^2} + \frac{2n + |l| + 1}{\sigma^2(t)}.
    \end{aligned}
\end{equation}
This system is very similar to the set of equations for the optical functions of a free electron state \eqref{OptEqFree}, yet it results in a drastically different dynamics.

Although one can take arbitrary initial conditions to specify the unique solution of the system \eqref{eq:Opt}, in a real experiment, they are determined by the incoming electron state. This prompts us to use the values of the dispersion, its time derivative, and the Gouy phase of the $\operatorname{NSLG}_{\text{f}}$ electron at the time $t_0$ when it enters the solenoid as the initial conditions for the $\operatorname{NSLG}_{\text{H}}$ state:
\begin{equation}
\label{InCond}
    \sigma(t_0) = \sigma_{\text{f}}(t_0) = \sigma_0, \hspace{30pt} \sigma'(t_0) = \sigma'_{\text{f}}(t_0) = \sigma'_0, \hspace{30pt} \Phi_{\text{G}}(t_0) = \Phi_{\text{f}}(t_0) = \Phi_0.
\end{equation}

Following the seminal approach of Silenko et al. \cite{Silenko2021}, we derive the dispersion of the $\operatorname{NSLG}_{\text{H}}$ electron from Eqs. \eqref{eq:Opt} with the initial conditions \eqref{InCond}:
\begin{equation}
\label{NSLGDisp}
    \begin{aligned}
        & \sigma(t) = \sigma_{\text{st}} \sqrt{1+ \sqrt{1 - \left( \frac{\sigma_{\text{L}}}{\sigma_{\text{st}}} \right)^4} \sin{ \left[ s(\sigma_0, \sigma_0') \omega (t-t_0) - \theta \right] }}, \\
        \sigma_{\text{st}}^2 = & \frac{\sigma_0^2}{2} \left( 1 + \left( \frac{\sigma_{\text{L}}}{\sigma_0} \right)^4 + \left( \frac{\sigma'_0 \sigma_{\text{L}}^2}{\lambda_{\text{C}} \sigma_0} \right)^2 \right), \hspace{30pt} \theta = \arcsin{\frac{1 - (\sigma_0/\sigma_{\text{st}})^2 }{\sqrt{1 - (\sigma_{\text{L}}/\sigma_{\text{st}})^4}}},
    \end{aligned}
\end{equation}
where the sign function is
\begin{equation}
    s(\sigma_0,\sigma_0') = \begin{cases}
     \operatorname{sgn}(\sigma_0'),\  \sigma_0' \ne 0,\\
     \operatorname{sgn}(\sigma_{\text{L}}-\sigma_0),\ \sigma_0'=0,\\
     0,\ \sigma_0 = \sigma_{\text{L}} \;\text{and}\; \sigma_0' = 0.
    \end{cases}
\end{equation}
This dispersion describes the oscillations of the r.m.s. radius of the electron inside the solenoid with a period $T_{\text{c}} = 2 \pi / \omega$. The value $\theta$ is the initial phase of the oscillations.

We should also note that states similar to those discussed in this section are presented in the books \cite{MalkinManko, BagrovGitman} as coherent states of an electron in the magnetic field with the vector potential \eqref{SymGauge}. Another approach to obtaining the $\operatorname{NSLG}_{\text{H}}$ wave functions using quantum Arnold transformation was recently realized in \cite{Filina2023}.

The parameter $\sigma_{\text{st}}^2$ is the period-averaged dispersion square
\begin{equation}
    \label{StDisp}
    \sigma_{\text{st}}^2 = \frac{1}{T_{\text{c}}} \int_0^{T_{\text{c}}} \sigma^2(t) dt \geq \sigma_{\text{L}}^2.
\end{equation}
We further use the corresponding time-averaged radius square
\begin{equation}
\label{StR}
    \rho^2_{\text{st}} = (2n + |l| +1 )\sigma^2_{\text{st}} \geq \rho^2_{\text{L}}
\end{equation}
as a characteristic size of the oscillating wave packet. The inequalities in Eqs. \eqref{StDisp}, \eqref{StR} are derived and discussed in Sec. \ref{secNSLGandLandauB}.

\begin{figure*}[b]
 \begin{subfigure}{0.49\textwidth}
     \includegraphics[width=\textwidth]{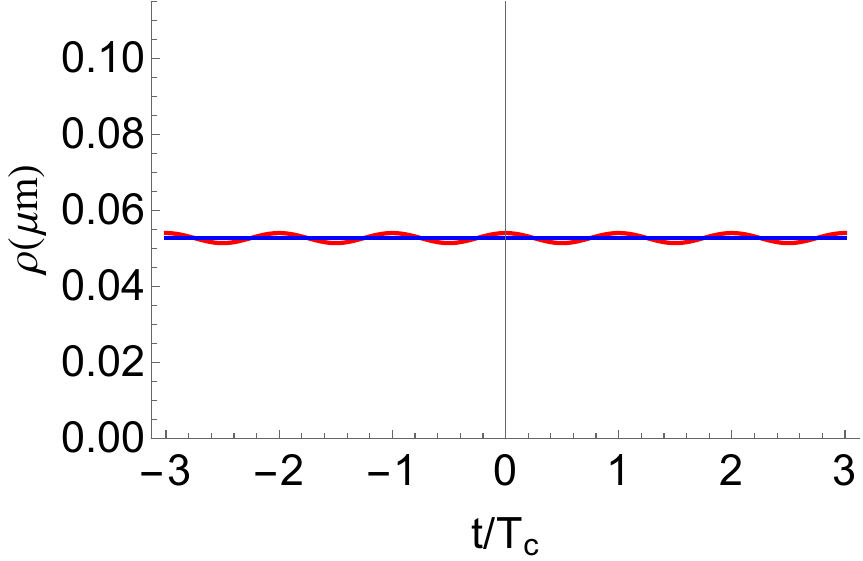}
     \caption{}
     \label{fig:NSLG_Similar}
 \end{subfigure}
 \hfill
 \begin{subfigure}{0.49\textwidth}
     \includegraphics[width=\textwidth]{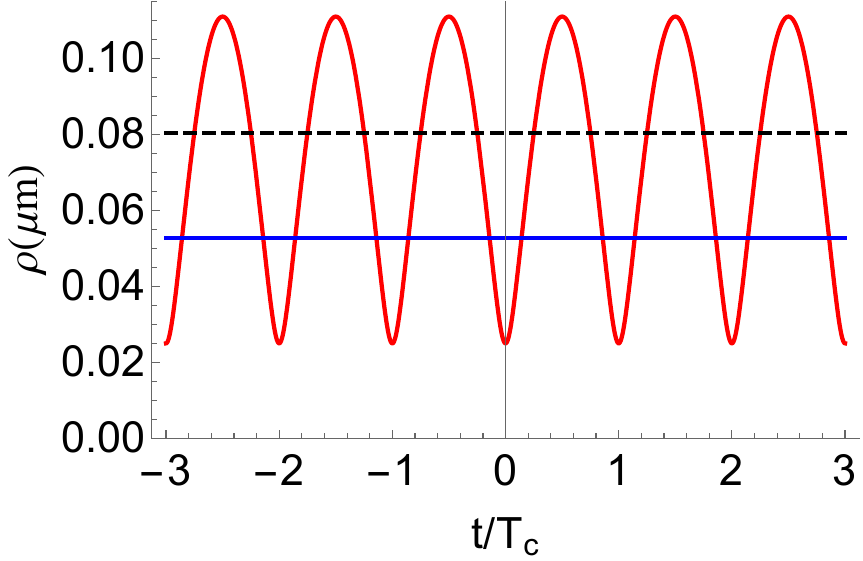}
     \caption{}
     \label{fig:NSLG_Less}
 \end{subfigure}
 
 \medskip
 \begin{subfigure}{0.49\textwidth}
     \includegraphics[width=\textwidth]{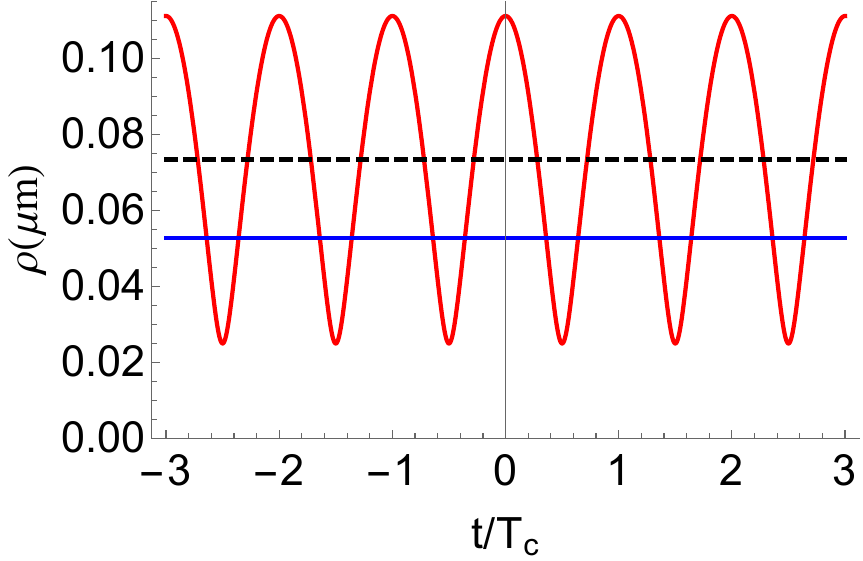}
     \caption{}
     \label{fig:NSLG_More}
 \end{subfigure}
 \hfill
 \begin{subfigure}{0.49\textwidth}
     \includegraphics[width=\textwidth]{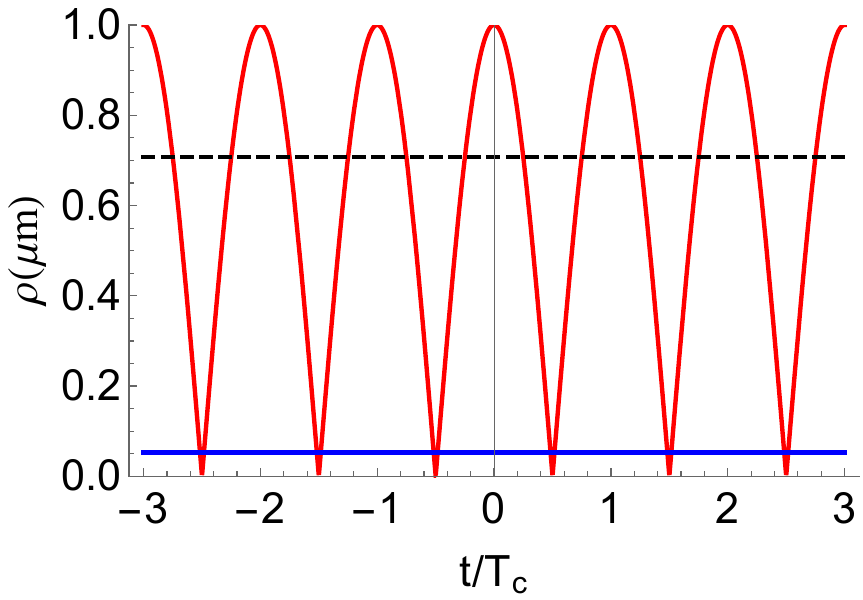}
     \caption{}
     \label{fig:NSLG_MuchMore}
 \end{subfigure}\caption{R.m.s. radius oscillations. The $\operatorname{NSLG}_{\text{H}}$ packet r.m.s. radii (in red) for different $\rho_0$; $\rho_{\text{L}} \approx 52.7$ nm is in blue. Black dashed lines correspond to $\rho_{\text{st}}$ given by Eq. \eqref{StR}. In each subfigure $H = 1.9$ T ($T_{\text{c}} \approx 0.02$ ns.), n = 0, l = 3, $\rho'_0 = 0$. (a)~$\rho_0 = 54$~nm, (b) $\rho_0 = 25$ nm, (c) $\rho_0 = 111.1$ nm, (d) $\rho_0 = 1 \mu$m.}
 \label{fig:InitialSizeInfluence}
\end{figure*}

The oscillations of the r.m.s.~radius of the $\operatorname{NSLG}_{\text{H}}$ states are shown in Fig. \eqref{fig:InitialSizeInfluence}. We consider the magnetic field $H = 1.9$ T, typical for transmission electron microscopes, and quantum numbers $n = 0$, $l = 3$ (the corresponding $\rho_{\text{L}} \approx 52.7$ nm) \cite{Schattschneider2014}. For simplicity, we set $\rho'_0 = 0$. A nonzero initial value of the divergence rate $\rho'_0$  alters the initial phase of the oscillations $\theta$ and the amplitude in accordance with Eqs.~\eqref{NSLGDisp}, but the picture remains qualitatively the same. We discuss how nonzero divergence rate affects the r.m.s. radius oscillations in the Appendix \ref{DivRateInf}.

Now let us discuss the possible oscillation regimes. In Fig. \ref{fig:NSLG_Similar}, the free electron size at the boundary $\rho_0 = 54$~nm is close to $\rho_{\text{L}}$. The r.m.s.~radius of the corresponding $\operatorname{NSLG}_{\text{H}}$ state oscillates around approximately the same value with a negligibly small amplitude. As we will discuss later (see Sec. \ref{secNSLGandLandauC}), such an electron can be considered to be in a Landau state to a good extent. In Fig. \ref{fig:NSLG_Less}, $\rho_0 = 25$ nm is significantly smaller than $\rho_{\text{L}}$. In this case, the magnetic field 
forces the wave packet to expand
to the size of the corresponding Landau state. By the time it happens, the r.m.s.~radius of the $\operatorname{NSLG}_{\text{H}}$ state acquires a nonzero divergence rate and continues broadening  past $\rho_{\text{L}}$. In Fig. \ref{fig:NSLG_More}, $\rho_0 = 111.1$ nm is larger than $\rho_{\text{L}}$, and their ratio is exactly the inverse of that in \ref{fig:NSLG_Less}. Here, in contrast, the field 
causes shrinking of the packet at first; as a result, the r.m.s.~radius decreases past the Landau state value and oscillates. Note that for two states with initial sizes $\rho_{0,1}$ and $\rho_{0,2}$, if $\rho_{0,1}/\rho_{\text{L}} = \rho_{\text{L}}/\rho_{0,2}$, the oscillations only differ by a $\pi$ phase shift and are otherwise identical. Finally, in Fig. \ref{fig:NSLG_MuchMore}, we consider an electron of the size $\rho_0 = 1 \, \mu$m much larger than $\rho_{\text{L}}$. Then, the oscillations of the r.m.s.~radius of the $\operatorname{NSLG}_{\text{H}}$ electron 
happen smoothly at larger values and the divergence rate changes abruptly at the lowest value of the r.m.s. radius. Similar behavior (shifted by half a period) is observed when the initial $\operatorname{NSLG}_{\text{H}}$ packet size is much less than the Landau radius.

Thus, from Fig.~\ref{fig:InitialSizeInfluence}, we can identify three oscillation regimes: 
\vspace{8pt}
\begin{enumerate}
    \item Landau-like regime: the r.m.s.~radius of the $\operatorname{NSLG}_{\text{H}}$ state is almost constant,
    \item Sine-like regime: the stationary r.m.s.~radius \eqref{StR} is always larger than the Landau radius, but they have the same order of magnitude,
    \item Bouncing regime: the r.m.s.~radius of the $\operatorname{NSLG}_{\text{H}}$ state 
    rapidly changes at its minimum, and 
    the time-averaged value of the r.m.s. radius is much larger than that of the Landau state.
\end{enumerate}

The oscillating behavior of the $\operatorname{NSLG}_{\text{H}}$ states' r.m.s. radius reminds that of optical Gaussian beams in ducts or graded-index optical waveguides \cite{Siegman}. A duct analogue of $\sigma_{\text{L}}^{-2}$ is $\sigma_{\text{O}}^{-2} = \lambda / (\pi \sqrt{n_2})$, where $\lambda$ is the beam wavelength in a medium and $n_2 = d^2 n(\rho) / d \rho^2 |_{\rho = 0}$ is the second derivative of the refractive index with respect to the radial coordinate near the symmetry axis. Now let us consider an optical Gaussian beam with a waist dispersion distinct form $\sigma_{\text{O}}$. In this case, the r.m.s.~radius of such a beam will oscillate similar to the r.m.s. radius of the $\operatorname{NSLG}_{\text{H}}$ state, whose oscillations are shown in Fig.~\ref{fig:InitialSizeInfluence}.

\begin{figure}[t]
\center{\includegraphics[width=0.8\linewidth]{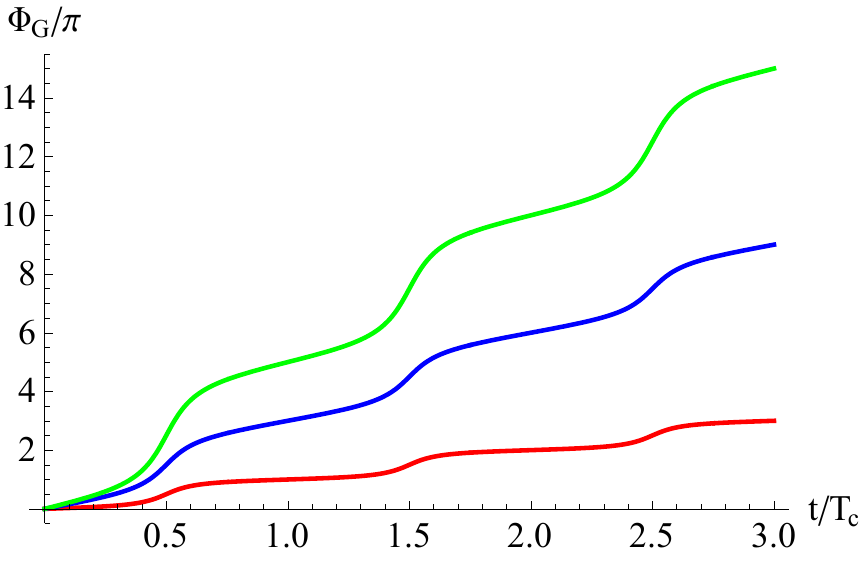}}
\caption{Gouy phase $\Phi_{\text{G}}(t)$ of the $\operatorname{NSLG}_{\text{H}}$ wave packet. The field strength $H = 1.9$ T (corresponding $\rho_{\text{L}} \approx 37$ nm), $\rho_0 \approx 71$ nm, $\rho'_0 = 0$, $T_{\text{c}} \approx 0.02$ ns, and $\Phi_0 = 0$. The quantum numbers are: $n = 0$, $l = 0$ (red line); $n = 0$, $l = 1$ (blue); and $n = 1$, $l = 1$ (green).}
\label{fig:GouyPhase}
\end{figure}

The Gouy phase of the $\operatorname{NSLG}_{\text{H}}$ state is
\begin{equation}
\label{GouyField}
\begin{aligned}
    \displaystyle & \Phi_{\text{G}}(t) = \Phi_0 + \frac{l \omega (t - t_0)}{2} + (2n + |l| + 1) s(\sigma_0, \sigma_0') \\
    & \hspace{-40pt} \times \Biggl[ \arctan \Biggl( \frac{\sigma^2_{\text{st}}}{\sigma_{\text{L}}^2} \tan{\frac{s(\sigma_0, \sigma_0') \omega (t - t_0) + \theta}{2} + \frac{\sigma^2_{\text{st}}}{\sigma_{\text{L}}^2}\sqrt{1 - \left( \frac{\sigma_{\text{L}}}{\sigma_{\text{st}}} \right)^4}} \Biggr) \Biggr. \\
    & \hspace{15pt} - \Biggl. \arctan \Biggl( \frac{\sigma^2_{\text{st}}}{\sigma_{\text{L}}^2} \tan{\frac{\theta}{2} + \frac{\sigma^2_{\text{st}}}{\sigma_{\text{L}}^2} \sqrt{1 - \left( \frac{\sigma_{\text{L}}}{\sigma_{\text{st}}} \right)^4}}\Biggr) \Biggr] .
\end{aligned}
\end{equation}
In Eq. \eqref{GouyField}, the arc tangent should be treated as a multivalued function for the Gouy phase to be continuous. The Gouy phase for $H = 1.9$ T ($\rho_{\text{L}} \approx 64$ nm), $\rho_0 \approx 122$ nm, $\rho'_0 = 0$, and $\Phi_0 = 0$ is shown in Fig. \ref{fig:GouyPhase}. The red, blue, and green lines correspond to three different pairs of quantum numbers $(n, l) = \{ (0, 0), (0, 1), (1, 1) \}$, respectively.

A free Gaussian beam gains a phase factor of $\pi$ while travelling from distant past to distant future \cite{Bliokh2012, Guzzinati2013, Schachinger2015, Bliokh2017, Feng2001, Siegman, Silenko2021}. Most of the phase gain is accumulated around the waist of the packet. A free Laguerre-Gaussian beam acquires a phase factor of $(2n + \abs{l} + 1) \pi$ the same way, propagating near its waist. Inside the field, the dynamics are periodic, and the electron state acquires this phase factor each cyclotron period. Moreover, interaction of the OAM with the field provides an additional Zeeman-type phase $l \pi$ \cite{Bliokh2012, Guzzinati2013}. Thus, the phase accumulated by the $\operatorname{NSLG}_{\text{H}}$ state per $T_{\text{c}}$ is $(2n + |l| + l + 1) \pi$.

\begin{figure}[t]
\center{\includegraphics[width=0.8\linewidth]{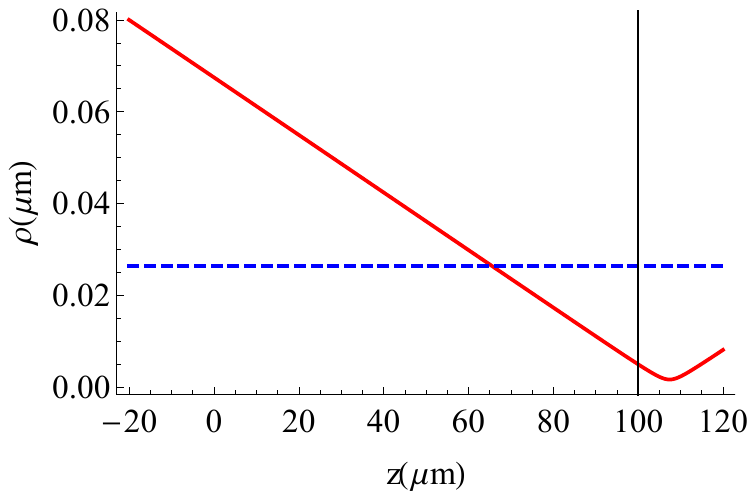}}
\caption{$\operatorname{NSLG}_{\text{H}}$ packet r.m.s. radius (in red) and Landau dispersion (in blue) based on the experiment \cite{Schattschneider2014}. We plot $\sigma_{\text{L}}$ and not $\rho_{\text{L}}$ to reproduce the Fig. 2b from \cite{Schattschneider2014}. A black vertical line marks the maximal value of $z$ in the work \cite{Schattschneider2014}. Longitudinal energy $E_{\parallel} = 200 $ KeV (corresponding velocity $v \approx 0.7 c$), $n = 0, l = 1$, $\rho_0 \approx 67.5$ nm, $\rho'_0 \approx -4.4 \times 10^{-4}$, and $H = 1.9$ T (corresponding $\sigma_{\text{L}} \approx 26$ nm).}
\label{fig:NSLGSchat}
\end{figure}

The average energy of the $\operatorname{NSLG}_{\text{H}}$ electron is
\begin{equation}
\label{NSLGEn}
    \expv{E} = \frac{\omega}{2} (2n + \abs{l} + 1) \frac{\sigma^2_{\text{st}}}{\sigma^2_{\text{L}}} + l \mu_{\text{B}} H .
\end{equation}
Generally, when $\sigma^2_{\text{st}} / \sigma_{\text{L}}^2 > 1$, the kinetic rotation prevails over the magnetic one. Moreover, for OAM directed opposite to the field, the two terms do not compensate each other, which removes the degeneracy of energy levels compared to the Landau states. Note that the average energy of the $\operatorname{NSLG}_{\text{H}}$ state \eqref{NSLGEn} is always larger than that of the Landau one \eqref{LandauEn}, and they are equal only for $\sigma_{\text{st}} = \sigma_{\text{L}}$, when the two states coincide (see Sec. \ref{secNSLGandLandau}).

Although the $\operatorname{NSLG}_{\text{H}}$ states have not yet been observed directly, an indirect evidence for their existence could have been obtained in the experiment of Schattschnider et al. \cite{Schattschneider2014}. In this experiment, the authors observed a possible part of the oscillations inherent to the $\operatorname{NSLG}_{\text{H}}$ states (see Fig. 2b in \cite{Schattschneider2014}). In Fig. \ref{fig:NSLGSchat}, we reproduce the evolution of the electron r.m.s. radius with the parameters from this work: electron energy $E_{\parallel} = 200$ KeV (corresponding velocity $v \approx 0.7 c$), $n = 0$, $l = 1$ (in the work, $l$ is designated as $m$), $\rho_0 \approx 67.5$ nm, $\rho'_0 \approx -4.4 \times 10^{-4}$, and $H = 1.9$ T. The black vertical line in Fig. \ref{fig:NSLGSchat} cuts off the $z$-region observed in the experiment. We extend this region a little to show the reader the subsequent 
growth in the r.m.s.~radius. Thus, we put forward the idea that the authors might have dealt with the $\operatorname{NSLG}_{\text{H}}$ state.

\section{Transversely relativistic wave packets}
\label{secRelativism}

We assume $E \ll mc^2$ while investigating twisted electrons in this work, but Eqs.~\eqref{FLGEn}, \eqref{LandauEn}, and \eqref{NSLGEn} make it clear that this condition is no longer valid for large $n$ and $\abs{l}$. Although in modern experiments, $n \sim 1$, beams with OAM values of several hundred \cite{Grillo2015} and even thousand $\hbar$ \cite{Mafakheri2017, McMorran2017} have already been generated. The restriction $E \ll mc^2$ sets the validity limits of our calculations and gives estimates of the quantum numbers that require relativistic treatment of the transverse dynamics. Furthermore, it allows considering beams that are transversely relativistic and longitudinally nonrelativistic, in contrast to those produced in accelerators nowadays.

Let us now estimate the quantum numbers $n$ and $l$ such that $\expv{E} \sim m c^2$. We start with an $\operatorname{NSLG}_{\text{f}}$ electron with energy $\expv{E}$ given by Eq. \eqref{FLGEnEx}. Using $\tau_\text{d} = \rho_\text{w}^2 / \left[ (2n + \abs{l} + 1) \lambda_\text{C} \right]$, we obtain a restriction on the quantum numbers of the free electron:
\begin{equation}
\label{FLGCond}
    2n + |l| + 1 \ll \frac{\rho_\text{w}}{\lambda_\text{C}}.
\end{equation}
Typically, twisted electrons are generated with $\rho_\text{w} \sim~1$~$\mu$m. For such particles, the value in the r.h.s.~of Eq.~\eqref{FLGCond} is of the order of $10^6$. However, being refocused to a 1 nm waist size, electrons with quantum numbers of the order of $10^3$ become transversely relativistic. Such focusing is easily achievable with appropriate magnetic lenses \cite{Schattschneider2014}. Thus, transversely relativistic free twisted electrons can be obtained in experiment already.

Applying the condition $\expv{E} \ll mc^2$ for a Landau state, we get
\begin{equation}
\label{LandauCond}
    \sqrt{(2n + \abs{l} + l + 1)} \ll \frac{\sigma_{\text{L}}}{\lambda_{\text{C}}}.
\end{equation}
Note that in free space, we fix the r.m.s.~radius of the generated electron $\rho_\text{w}$, but in a magnetic field, it is the dispersion $\sigma_{\text{L}}$ that is defined by the field strength. For example, if the field strength is of the order of 1 T, $\sigma_{\text{L}} \sim 36$ nm, and the r.h.s.~of the inequality \eqref{LandauCond} is of the order of $10^5$. For negative values of $l$, the l.h.s.~of Eq.~\eqref{LandauCond} does not depend on OAM at all. Therefore, when the magnetic and the kinetic rotations of the Landau state compensate each other, such a state remains nonrelativistic for any attainable values of $n$ and $\abs{l}$. However, for $l > 0$, the relativistic regime cannot be achieved either, as it would require OAM of the order of $10^{10}$.

For the $\operatorname{NSLG}_{\text{H}}$ states, the relativistic regime is more feasible than for the Landau counterparts, because
$\operatorname{NSLG}_{\text{H}}$ kinetic energy is enhanced by the factor $\sigma^2_{\text{st}}/\sigma^2_{\text{L}}$. Indeed, for an $\operatorname{NSLG}_{\text{H}}$ wave packet, we obtain \begin{equation}
    \sqrt{ \left[ \left( 2n + \abs{l} + 1 \right) \frac{\sigma_{\text{st}}^2}{\sigma_{\text{L}}^2} + l \right]} \ll \frac{\sigma_{\text{L}}}{\lambda_{\text{C}}}.
\end{equation}
Usually, the factor $\sigma_{\text{st}}^2 / \sigma_{\text{L}}^2 \gg 1$; for example, in the work \cite{Schattschneider2014}, $\sigma_{\text{st}}^2 / \sigma_{\text{L}}^2 \approx 31$. This allows us to simplify the above condition:
\begin{equation}
\label{NSLGNonRelCond}
    \sqrt{\left( 2n + \abs{l} + 1 \right)} \ll \frac{\sigma_{\text{L}}}{\lambda_{\text{C}}} \frac{\sigma_{\text{L}}}{\sigma_{\text{st}}}.
\end{equation}

The additional factor $\sigma_{\text{L}} / \sigma_{\text{st}}$  in the r.h.s.~of this inequality eases the requirements on the quantum numbers to obtain transversely relativistic states. For instance, in the experiment of Schattschnider and colleagues \cite{Schattschneider2014}, the r.h.s.~of Eq. \eqref{NSLGNonRelCond} is of the order of $10^4$. This value can be reduced even more, for example, by increasing $\rho_0$. To increase $\rho_0$, one can simply move the solenoid further from the source of twisted electrons. For large wave packets with $\sigma_0 \gg \sigma_{\text{L}}$ and with a sufficiently low divergence rate $\sigma'_0 \ll \lambda_{\text{C}} / \sigma_0$, the condition \eqref{NSLGNonRelCond} turns into
\begin{equation}
\label{NSLGNonRelCondSimp}
    \sqrt{\left( 2n + \abs{l} + 1 \right)} \ll \frac{\sigma_{\text{L}}}{\lambda_{\text{C}}} \frac{\sigma_{\text{L}}}{\sigma_0}.
\end{equation}
From here it follows that for wave packets with $\sigma_0 / \sigma_{\text{L}} \geq \sigma_{\text{L}} / \lambda_{\text{C}}$, even a Gaussian mode with $n = l = 0$ is relativistic. For a field strength of the order of 1 T, this happens when $\sigma_0 \sim 1$ mm, which can also be decreased if the divergence rate $\sigma'_0$ in \eqref{NSLGNonRelCond} is taken into account.

\section{Connection between nonstationary Laguerre-Gaussian states in free space and in the field}
\label{secNSLGandFree}

Before considering $\operatorname{NSLG}_{\text{H}}$ states in detail, we should note that their explicit wave function was obtained from the continuity of the optical functions at the boundary~\eqref{InCond}. In reality, not only these functions, but also the wave function itself is continuous. This is not surprising, because electron states in free space and inside the solenoid are defined by the ansatz of the same general form~\eqref{NSLG}.

We also need to make a special note about the energies of the $\operatorname{NSLG}_{\text{f}}$ and $\operatorname{NSLG}_{\text{H}}$ states. Generally, the quantities given by Eqs. \eqref{FLGEnEx} and \eqref{NSLGEn} are not equal to each other, i.e.~the energy is discontinuous at the boundary. This is a result of the energy dispersion, as the continuity of the average kinetic momentum $\expv{\hat{\bm{p}}}$ does not provide that of $\expv{E} \sim \expv{\hat{\bm{p}}^2} \ne \expv{\hat{\bm{p}}}^2$.

\subsection{Vanishing magnetic field}
\begin{figure}[t]
\center{\includegraphics[width=0.8\linewidth]{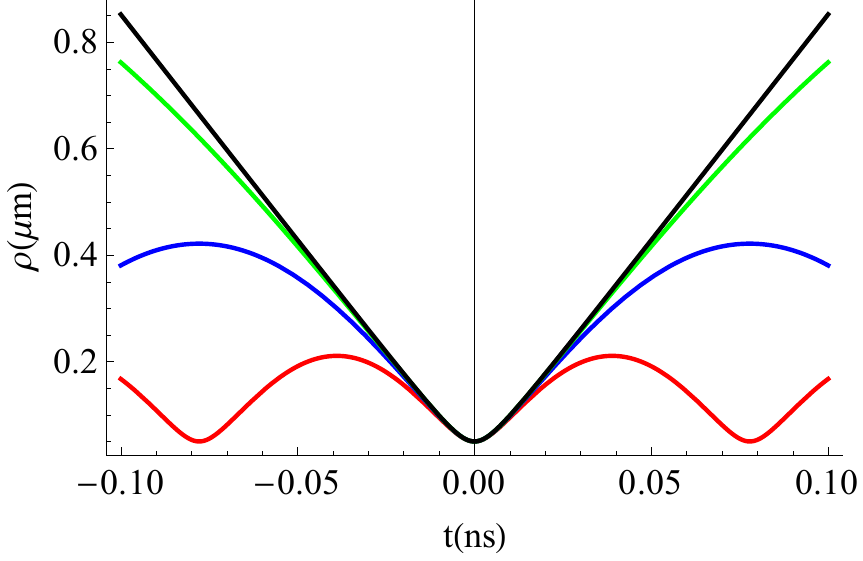}}
\caption{Vanishing magnetic field limit. R.m.s.~radius of an $\operatorname{NSLG}_{\text{H}}$ wave packet tends to $\rho_{\text{f}}(t)$ as the magnetic field decreases. $\operatorname{NSLG}_{\text{H}}$ packet r.m.s.~radii are shown for: $H = 0.5$ T (in red), $H = 0.25$ (in blue), $H = 0.1$ T (in green); The $\operatorname{NSLG}_{\text{f}}$ electron r.m.s.~radius $\rho_{\text{f}}(t)$ is indicated with the black line. 
The following parameters are used: 
$t_{\text{g}} = t_0 = 0$, $\rho_0 = \rho_{\text{w}} = 50$ nm, $\rho'_0 = \rho'_{\text{w}} = 0$, n = 0, and l = 3.}
\label{fig:FieldtoFree}
\end{figure}

One of the advantages of the $\operatorname{NSLG}_{\text{H}}$ states compared to the Landau ones is, they smoothly transform into free twisted electron wave packets in the vanishing magnetic field limit. To confirm this, we can find the limit of $\sigma(t), \Phi_{\text{G}}(t)$ as $H \rightarrow 0$ ($\sigma_{\text{L}} \rightarrow \infty$), see Appendix \ref{Limit} for rigorous derivation. In Fig.~\ref{fig:FieldtoFree}, we show how $\operatorname{NSLG}_{\text{H}}$ dispersion transforms into that of the $\operatorname{NSLG}_{\text{f}}$ as the magnetic field goes to zero.

In contrast, the Landau states dispersion diverges in the vanishing magnetic field limit, and the wave functions become delocalized.

\subsection{Off-axis injection}

\begin{figure}
\center{\includegraphics[width=0.8\linewidth]{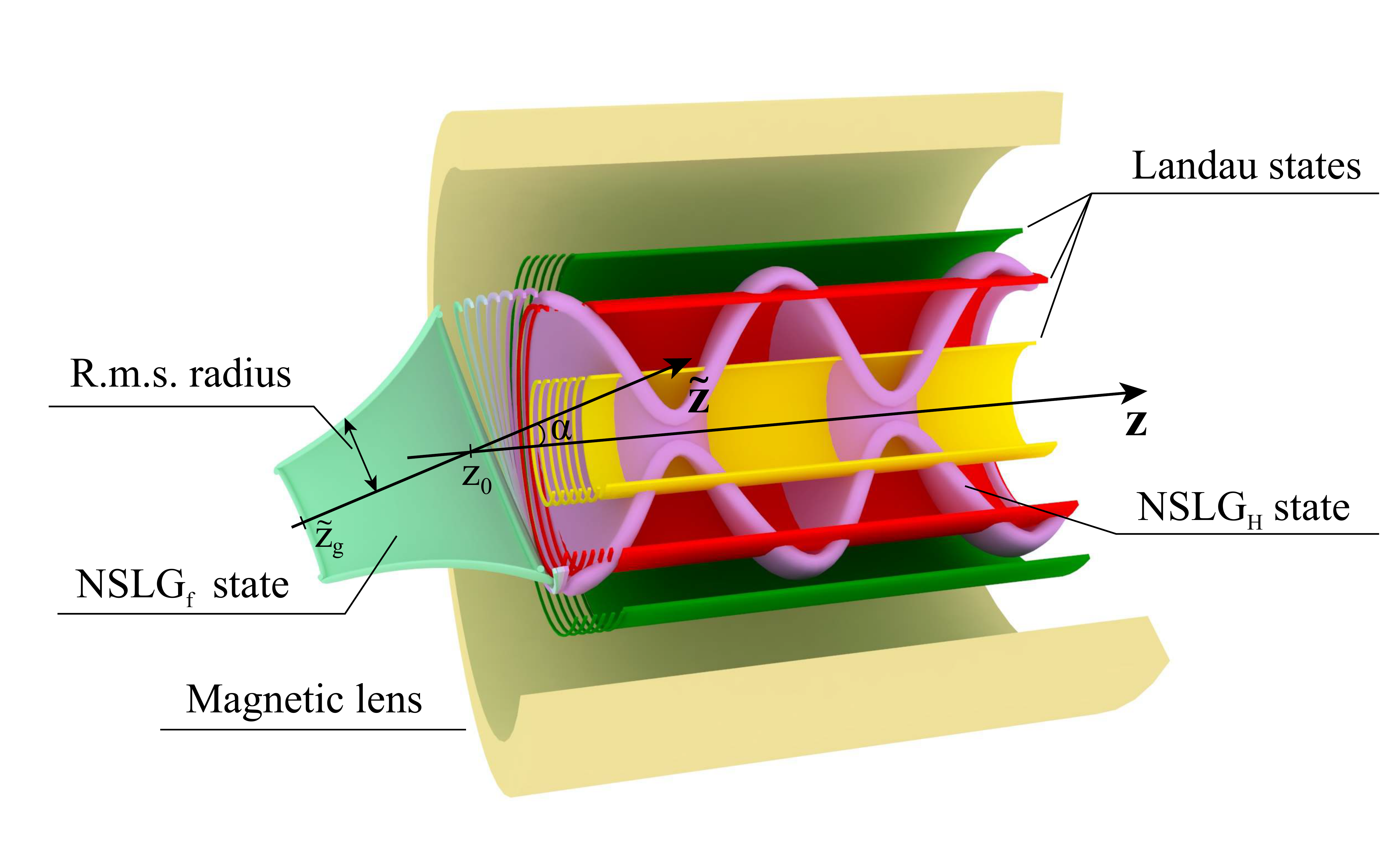}}
\caption{Free twisted electron entering a magnetic lens at a small angle $\alpha$ with respect to the field direction.}
\label{fig:LensOffAxis}
\end{figure}

In a real-life setup, the propagation axis of a twisted electron wave packet cannot be perfectly aligned with the magnetic field direction. Such a misalignment can be caused by a shift of the electron source or slight inhomogeneities of the magnetic field inside the solenoid. In this section, we account for this inaccuracy by considering a twisted electron that enters the lens at a small angle $\alpha$ with respect to the $z$-axis, as shown in Fig. \ref{fig:LensOffAxis}.

Imagine that by the time $t_0$ a free electron reaches the lens boundary at $z_0$, the propagation axis of the electron is shifted by the angle $\alpha$ with respect to the $z$-axis aligned with the field. The wave function of the corresponding state is given by
\begin{equation}
\label{rot}
    \tilde{\Psi}_{n l}(\bm{r}, t) = \Psi_{n l}(\tilde{\bm{r}}, t) = \Psi_{n l}(\tilde{\bm{\rho}},t)\Psi_{\parallel}(\tilde{z}, t),
\end{equation}
where the tilted coordinates 
\begin{equation}
\label{tildedco}
\begin{aligned}
    & \tilde{z} = z_0 -\rho\cos\varphi\sin\alpha + (z-z_0)\cos\alpha, \\
    & \tilde{\rho} = \sqrt{(\rho\cos\varphi\cos\alpha+(z-z_0)\sin\alpha)^2+\rho^2\sin^2\varphi},\\
    & \tilde{\varphi} = \arctan\left(\frac{\sin\varphi}{\cos\varphi\cos\alpha+\frac{z-z_0}{\rho}\sin\alpha}\right)
\end{aligned}
\end{equation}
are obtained by a rotation around the axis indicated by $\varphi = \pi / 2$. The rotational symmetry of the problem enables an arbitrary choice of the rotation axis in the transverse plane without any influence on the results. The transverse and longitudinal parts of the wave function in Eq. \eqref{rot} are given by Eqs. \eqref{NSLG} and \eqref{lo}, respectively. 

Let us now decompose the rotated wave function in terms of the electron states propagating along the $z$-axis:
\begin{equation}
\label{deco}
    \tilde{\Psi}_{n l}(\bm{r},t) = \sum\limits_{n', l'} \int\limits_{-\infty}^{\infty}\frac{dp'_z}{2\pi} c_{n n' l l'}(p'_z) \Psi_{n' l'}(\bm{\rho},t) g(p'_z) \exp(i p'_z z - i \frac{p_z^{\prime 2}}{2 m} t).
\end{equation}
Here, the decomposition coefficients are
\begin{equation}
\label{o}
\begin{aligned}
    & c_{n n' l l'}(p'_z) = \int d^2 \rho dz \Psi_{n' l'}^{*}(\bm{\rho},t) \Psi_{n  l}(\tilde{\bm{\rho}}, t) \\
    \times \int\limits_{-\infty}^{\infty}\frac{dp_z}{2\pi} & \exp(- i p'_z z + i \frac{p^{\prime 2}_z}{2m} t) \frac{g(p_z)}{g(p_z')} \exp(i p_z \tilde{z} - i \frac{p_z^2}{2m} t).
\end{aligned}
\end{equation}
We are interested in the off-axis corrections to the electron state in the vicinity of the lens boundary. Therefore, we evaluate $\tilde{\Psi}_{n l}(\bm{r}, t)$ at $z = z_0$ and $t = t_0$ in Eq.~\eqref{deco}. 

In the first non-vanishing order in $\alpha$ and for $z = z_0$, Eqs.~\eqref{tildedco} are simplified to
\begin{equation}
     \tilde{\rho} = \rho + o(\alpha^2), \hspace{30pt} \tilde{\varphi} = \varphi + o(\alpha^2), \hspace{30pt} \tilde{z} = z - \alpha \rho \cos{\varphi} + o(\alpha^2),
\end{equation}
and the coefficients \eqref{o} take the form
\begin{equation}
   c_{n n' l l'}(p'_z) = \int \Psi^*_{n' l'}(\bm{\rho}, t)\Psi_{n l}(\bm{\rho}, t) \exp(-i\alpha p'_z \rho \cos\varphi)d^2\rho.
\end{equation}

The integral over the transverse plane can be evaluated using Eq. (7.422) in \cite{Gradshtein} (there is, however, a misprint $m\leftrightarrow n$ in the book). The absolute value of the coefficients is
\begin{equation}
\label{coef}
\begin{aligned}
|c_{n n' l l'}|(p'_z) = & \delta_{n, n'} \delta_{l, l'} + \frac{\alpha p'_z \sigma(t_0)}{4\pi} \delta_{|l'|, |l|-1} \left[\delta_{n', n} \sqrt{n + |l|} + \delta_{n', n + 1} \sqrt{n + 1} \right] \\
& + \frac{\alpha p'_z \sigma(t_0)}{4\pi} \delta_{|l'|, |l| + 1} \left[\delta_{n', n} \sqrt{n + |l| + 1} + \delta_{n', n - 1} \sqrt{n} \right]. 
\end{aligned}
\end{equation}

If the longitudinal wave functions have a sufficiently narrow distribution in coordinate and momentum spaces simultaneously, we can evaluate the decomposition coefficients in a different manner. First, we can approximate the integrals over the longitudinal momentum by evaluating the integrand at the mean value $p_z = \expv{p_z}$. Then, Eq. \eqref{deco} becomes
\begin{equation}
\label{decotoo}
    \Psi_{n l}(\tilde{\bm{\rho}}, t) \exp(-i \alpha \expv{p_z} \rho \cos\varphi) = \sum\limits_{n',l'} c_{n n' l l'}(\expv{p_z}) \Psi_{n' l'}(\bm{\rho},t).
\end{equation}
The expression \eqref{decotoo}, as compared to Eq. \eqref{deco}, does not contain the longitudinal wave function, whose entire contribution is accounted for by the average momentum $\expv{p_z}$. Proceeding in the same manner, we get
\begin{equation}
\label{coef1}
\begin{aligned}
|c_{n n' l l'}| = & \delta_{n, n'}\delta_{l, l'} + \frac{\alpha \expv{p_z} \sigma(t_0)}{4\pi} \delta_{|l'|, |l|-1} \left[ \delta_{n', n} \sqrt{n + |l|} + \delta_{n', n+1} \sqrt{n + 1} \right] \\
& + \frac{\alpha \expv{p_z} \sigma(t_0)}{4\pi} \delta_{|l'|, |l| + 1} \left[ \delta_{n', n} \sqrt{n + |l| + 1} + \delta_{n', n - 1} \sqrt{n} \right]. 
\end{aligned}
\end{equation}

From Eq. \eqref{coef1}, we see that the actual dimensionless parameter defining the magnitude of the coefficients is $\alpha\expv{p_z}\sigma(t_0)$. In real life, the value of $\sigma(t_0)$ is of the order of several $\mu$m or less. Provided that currently $n \sim 1$ , $l \lesssim 10^4$, even for 10 GeV-electrons with $\expv{p_z} \sim 10^{-3} \mu \text{m}^{-1}$, we obtain $|c_{n n' l l'}| \lesssim 10^{-2} \alpha$. This means that the off-axis corrections are negligible for any feasible experimental scenario.

\section{Connection between nonstationary Laguerre-Gaussian states in solenoid and Landau states}
\label{secNSLGandLandau}

\subsection{Landau states as a special case of nonstationary Laguerre-Gaussian states}
\label{secNSLGandLandauA}

Although the Landau states \eqref{Landau} are represented by stationary wave functions, they also have the form \eqref{NSLG}. Moreover, both $\operatorname{NSLG}_{\text{H}}$ and Landau states are solutions of the Schr\"{o}dinger equation \eqref{Schr} with the same Hamiltonian \eqref{HamField}, which leads to the same system of optical equations \eqref{eq:Opt}. Here, the question arises: how these two sets of states are linked?

To answer this question, one may look for a solution of the system \eqref{eq:Opt} corresponding to the stationary Landau states. Such a solution exists for the unique choice of the initial conditions:
\begin{equation}
\label{UniquInCond}
    \sigma_0 = \sigma_{\text{L}}, \hspace{30pt} \sigma_0' = 0.
\end{equation}
This means that the Landau states are but a special case of the $\operatorname{NSLG}_{\text{H}}$ ones forming when a free twisted electron with a specific size and zero divergence rate crosses the boundary. Otherwise, an electron inside the solenoid is described by general $\operatorname{NSLG}_{\text{H}}$ states rather than the Landau ones.

To characterize the deviation of the $\operatorname{NSLG}_{\text{H}}$ states from the Landau ones, we introduce two dimensionless parameters
\begin{equation}
\label{Nonlandauity}
    \xi_1 = \frac{\sigma_{\text{L}}}{\sigma_0}, \hspace{30pt} \xi_2 = \frac{\abs{\sigma_0'} \sigma_{\text{L}}}{\lambda_{\text{C}}}.
\end{equation}
From Eq. \eqref{UniquInCond}, it follows that for the Landau states $\xi_1 = 1$, $\xi_2 = 0$. The more these parameters differ from 1 and 0, respectively, the more distinguishable the $\operatorname{NSLG}_{\text{H}}$ and the Landau states are. This effect manifests itself most clearly in growing amplitude of the r.m.s.~radius oscillations and its period-averaged value.

\subsection{Comparison of sizes of nonstationary Laguerre-Gaussian states in solenoid and Landau states}
\label{secNSLGandLandauB}

To characterize the size of an $\operatorname{NSLG}_{\text{H}}$ electron, we use the stationary radius $\rho_{\text{st}}$ given by Eq. \eqref{StR}. Naively, it seems that this value should be equal to or at least close to $\rho_{\text{L}}$ \cite{Greenshields2014, Greenshields2015, Karlovets2021Vortex}. However, this is generally not true. In terms of the parameters \eqref{Nonlandauity}, $\rho_{\text{st}}$ is expressed as
\begin{equation}
\label{StRadNonl}
    \rho_{\text{st}} = \rho_{\text{L}} \left[ \frac{\xi_1^2 + \xi_1^{-2}}{2} + \frac{\xi_2^2}{2}\right]^{1/2} \ge \rho_{\text{L}}.
\end{equation}
From this expression, it is clear that for $\xi_1 \gg 1$, $\xi_1 \ll 1$, or $\xi_2 \gg 1$, the relation $\rho^2_{\text{st}} \gg \rho_{\text{L}}^2$ holds. In contrast, for the initial conditions \eqref{UniquInCond}, when the electron in the field is indeed in the Landau state, the minimum value $\rho_{\text{st}} = \rho_{\text{L}}$ is reached. This illustrates that boundary conditions significantly affect the electron states inside the lens.

The conditions imposed on the parameters $\xi_{1, 2}$ for the $\operatorname{NSLG}_{\text{H}}$ state to be close to a Landau one are very specific. Unless an experimenter is intended to obtain a Landau state, an $\operatorname{NSLG}_{\text{H}}$ state is almost certainly generated. For example, in the experiment of Schattschnider et al. \cite{Schattschneider2014}, the parameters of the setup $n = 0$, $\abs{l} = 1$, $\sigma_0 = 4.77 \times 10^{-2} \mu$m, and $\sigma'_0 = -3.1 \times 10^{-4}$ lead to $\xi_1 = 0.76$ and $\xi_2 = 29.21 \gg 1$. For these parameters, we find $\rho_{\text{st}} = 20.7 \rho_{\text{L}} \gg \rho_{\text{L}}$, which again supports our idea that $\operatorname{NSLG}_{\text{H}}$ states 
were observed in the work \cite{Schattschneider2014}.

\subsection{Decomposition of nonstationary Laguerre-Gaussian states in the field in terms of Landau ones}
\label{secNSLGandLandauC}

Comparing the characteristic sizes of an $\operatorname{NSLG}_{\text{H}}$ and a Landau state, we qualitatively estimate the difference between the two states. For a more substantive investigation, we should decompose an $\operatorname{NSLG}_{\text{H}}$ state wave function in terms of the stationary Landau ones \eqref{Landau}: 
\begin{equation}
\label{NSLGDecLandau}
\Psi_{n l}(\bm{\rho},t) = \sum_{n', l'} a_{n n' l} \delta_{l, l'} \Psi^{(\text{L})}_{n' l'}(\bm{\rho},t).
\end{equation}

Since the evolution of both sides in Eq. \eqref{NSLGDecLandau} is governed by the same Hamiltonian, the decomposition coefficients do not depend on time. We present the explicit expression for $a_{n n' l}$ in the Appendix \ref{NSLGDecLandauApp}. Note that the Kronecker delta reflects the OAM conservation.

As we have discussed in the previous section, $\rho_{\text{st}} = \rho_{\text{L}}$ only when $\operatorname{NSLG}_{\text{H}}$ and Landau states coincide. Indeed, from this equality, it follows that
\begin{equation}
\label{pm}
    a_{n n' l} = \delta_{n, n'}.
\end{equation}
However, in experiment, it is impossible to precisely satisfy the initial conditions \eqref{UniquInCond} to obtain a single Landau mode inside the solenoid.

Let us analyze what happens to the $\operatorname{NSLG}_{\text{H}}$ state inside the lens when its characteristic size and $\rho_{\text{L}}$ with the same quantum numbers $n, l$ are close, yet not equal:
\begin{equation}
    \delta \zeta = \frac{\rho_{\text{st}} - \rho_{\text{L}}}{\rho_{\text{L}}} \ll 1.
\end{equation}
This is true when the size of the incoming packet at the boundary slightly differs from $\rho_{\text{L}}$ and the divergence rate is low. From Eq. \eqref{NSLGDisp}, we know that in this situation, rather than being constant and equal to $\rho_{\text{L}}$, the r.m.s.~radius inside the lens begins oscillating around a slightly larger value, $\rho_{\text{st}}$, with a small amplitude. The decomposition coefficients clearly indicate that for a small detuning, a few neighbouring Landau modes contribute to the $\operatorname{NSLG}_{\text{H}}$ state:
\begin{equation}
\label{decs}
    a_{n n' l} \propto (\delta \zeta)^{\frac{{|n' - n|}}{2}}.
\end{equation}
Interference of these different states results in the r.m.s.~radius oscillations and a change in the period-averaged size.

\begin{figure*}[t]
 \begin{subfigure}{0.24\textwidth}
     \includegraphics[width=\textwidth]{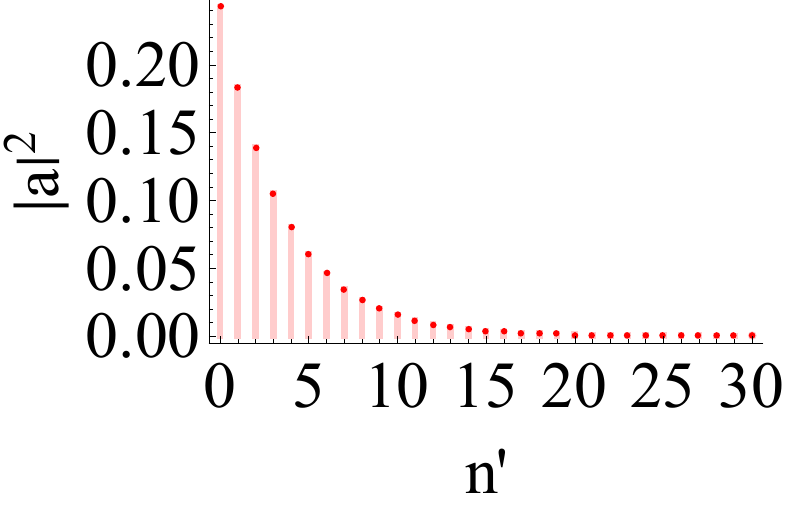}
     \caption{}
     \label{fig:Decomp0_0}
 \end{subfigure}
 \hfill
 \begin{subfigure}{0.24\textwidth}
     \includegraphics[width=\textwidth]{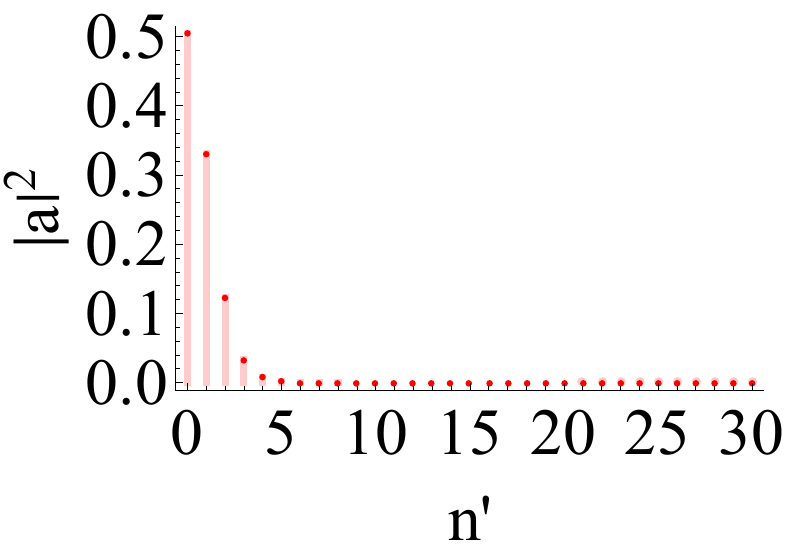}
     \caption{}
     \label{fig:Decomp0_7}
 \end{subfigure}
  \hfill
 \begin{subfigure}{0.24\textwidth}
     \includegraphics[width=\textwidth]{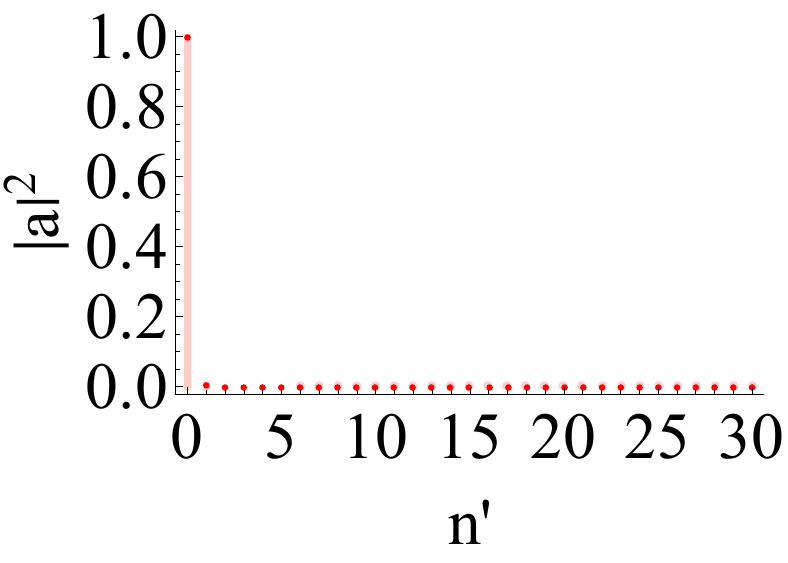}
     \caption{}
     \label{fig:Decomp0_13}
 \end{subfigure}
 \hfill
 \begin{subfigure}{0.24\textwidth}
     \includegraphics[width=\textwidth]{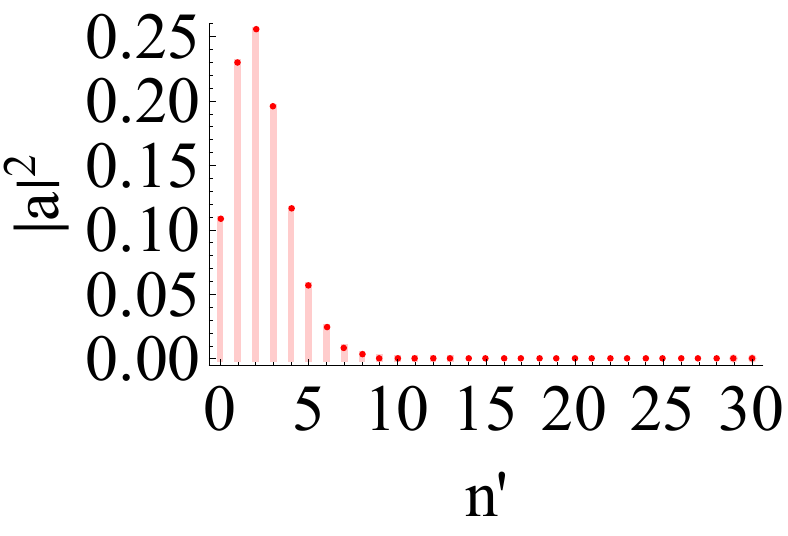}
     \caption{}
     \label{fig:Decomp0_25}
 \end{subfigure}
 
 \medskip
 \begin{subfigure}{0.24\textwidth}
     \includegraphics[width=\textwidth]{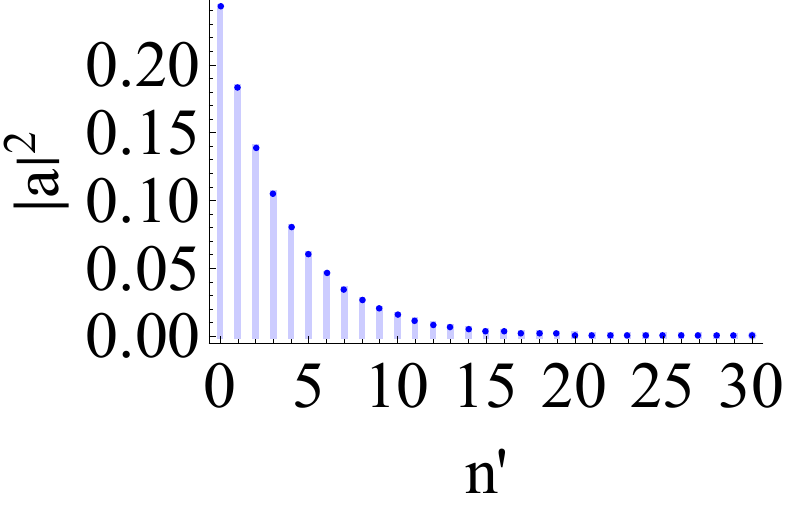}
     \caption{}
     \label{fig:Decomp0_0_}
 \end{subfigure}
 \hfill
 \begin{subfigure}{0.24\textwidth}
     \includegraphics[width=\textwidth]{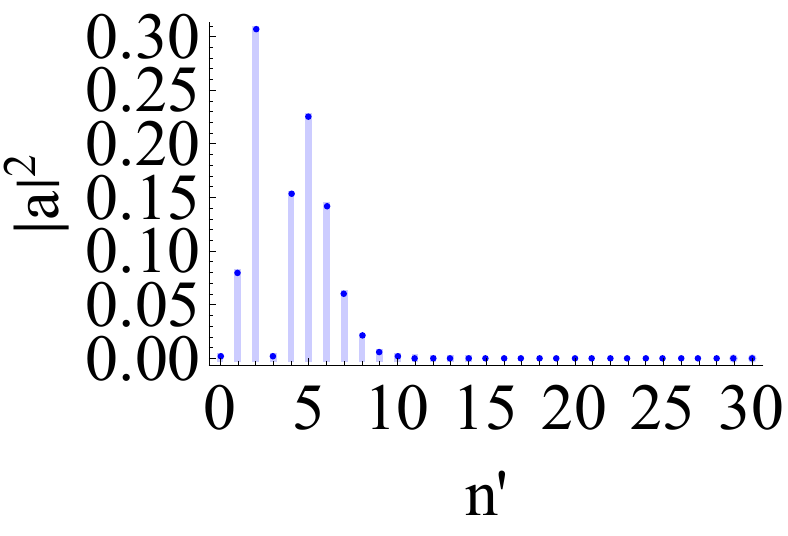}
     \caption{}
     \label{fig:Decomp3_0}
 \end{subfigure}
  \hfill
 \begin{subfigure}{0.24\textwidth}
     \includegraphics[width=\textwidth]{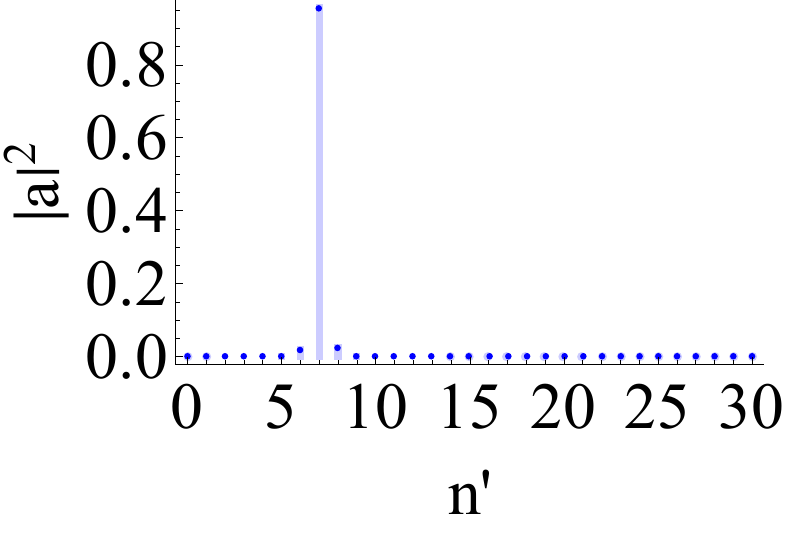}
     \caption{}
     \label{fig:Decomp7_0}
 \end{subfigure}
 \hfill
 \begin{subfigure}{0.24\textwidth}
     \includegraphics[width=\textwidth]{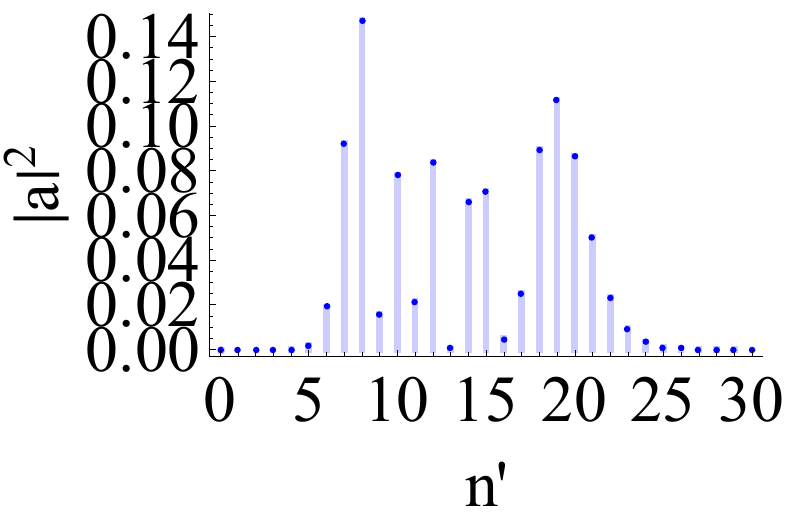}
     \caption{}
     \label{fig:Decomp12_0}
 \end{subfigure}

 \medskip
 \begin{subfigure}{0.24\textwidth}
     \includegraphics[width=\textwidth]{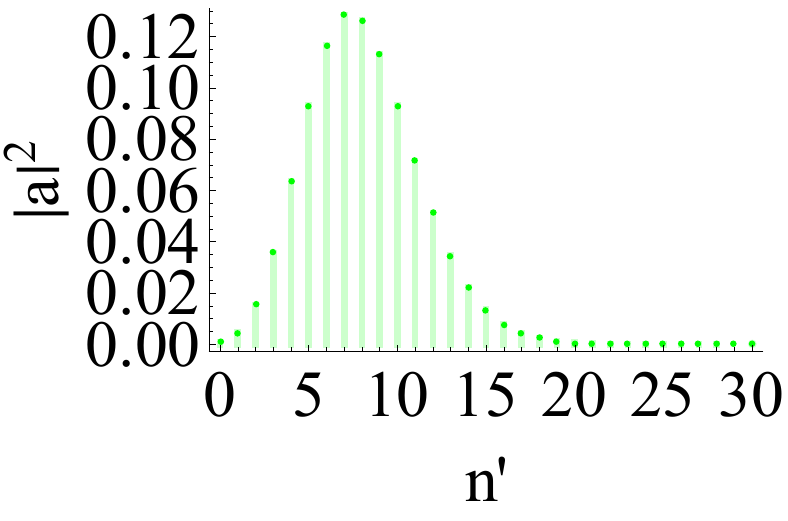}
     \caption{}
     \label{fig:Decomp0_35}
 \end{subfigure}
 \hfill
 \begin{subfigure}{0.24\textwidth}
     \includegraphics[width=\textwidth]{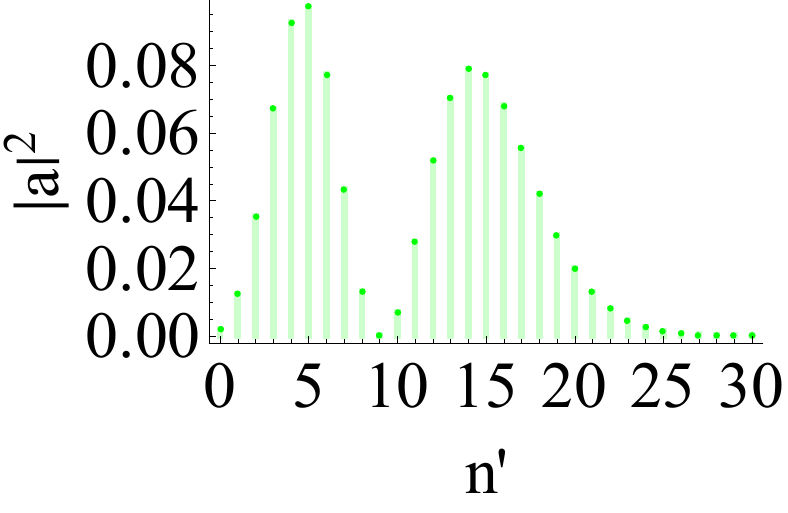}
     \caption{}
     \label{fig:Decomp1_35}
 \end{subfigure}
  \hfill
 \begin{subfigure}{0.24\textwidth}
     \includegraphics[width=\textwidth]{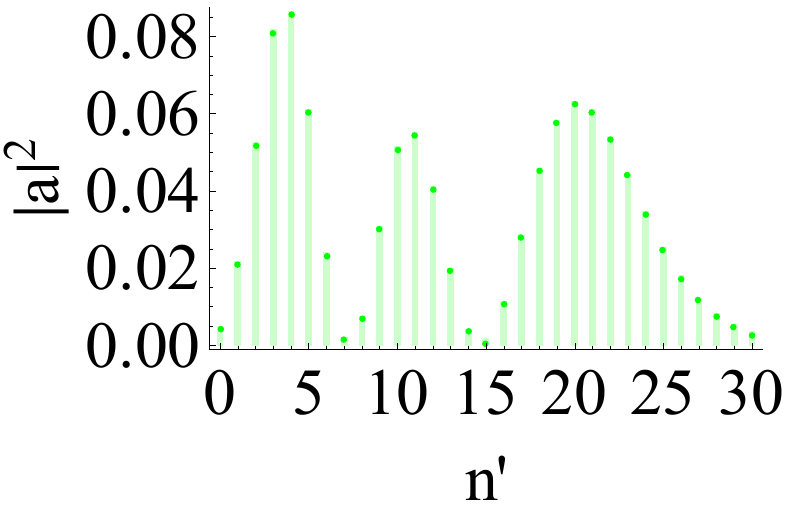}
     \caption{}
     \label{fig:Decomp2_35}
 \end{subfigure}
 \hfill
 \begin{subfigure}{0.24\textwidth}
     \includegraphics[width=\textwidth]{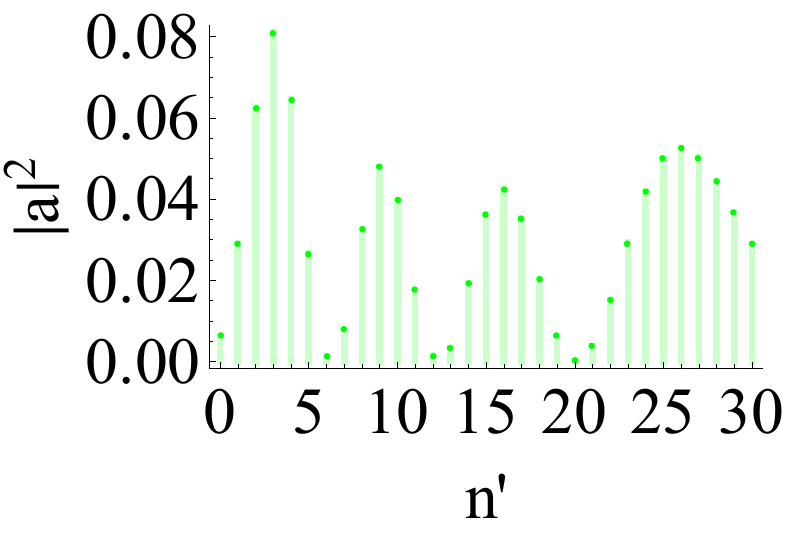}
     \caption{}
     \label{fig:Decomp3_35}
 \end{subfigure}

 \caption{Probability coefficients $\abs{a_{n n' l}}^2$  for an NSLG state with the quantum numbers $n$, $l$ decomposed into a superposition of the Landau states with the quantum numbers $n'$, $l$. $\rho_0 = 100$ nm, $\rho'_0 = 0$, $B = 1.9$ T. Top panel (in red): (a) $n = 0$, $l = 0$, (b) $n = 0$, $l = 7$, (c) $n = 0$, $l = 13$, (d) $n = 0$, $l = 25$; middle panel (in blue): (e) $n = 0$, $l = 0$, (f) $n = 3$, $l = 0$, (g) $n = 7$, $l = 0$, (h) $n = 12$, $l = 0$; bottom panel (in green): (i) $n = 0$, $l = 35$, (j) $n = 1$, $l = 35$, (k) $n = 2$, $l = 35$, (l) $n = 3$, $l = 35$.}
 \label{fig:Decomp}
\end{figure*}

A more complex picture arises when the state inside the solenoid significantly differs from any of the Landau states, i.e. $\xi_1 \gg 1$, or $\xi_1 \ll 1$, or $\xi_2 \gg 1$. In this case, the $\operatorname{NSLG}_{\text{H}}$ state is a superposition of numerous Landau ones. The coefficients form wide, oscillating distributions as functions of the radial quantum number of the Landau states $n'$. 

Examples of the probability coefficients $\abs{a_{n n' l}}^2$ for the possible scenarios are presented in Fig.~\ref{fig:Decomp}. We choose $\rho_0 = 100$ nm, $\rho'_0 = 0$, and $H = 1.9$ T, for which $\sigma_{\text{L}} \approx 26$ nm.

In Figs. \ref{fig:Decomp0_0} --- \ref{fig:Decomp0_25} (top panel, in red), we study the distribution of $\abs{a_{n  n'  l}}^2$ for different values of $l$ while keeping $n = 0$. In Fig. \ref{fig:Decomp0_0}, $l = 0$ and $\rho_{\text{L}} \approx 26$ nm, so the $\operatorname{NSLG}_{\text{H}}$ state is wider than the Landau one with corresponding quantum numbers. As a consequence, higher-order Landau modes appear in the decomposition. Then, with increasing OAM (Figs. \ref{fig:Decomp0_7}, \ref{fig:Decomp0_13}), $\rho_{\text{L}}$ gets closer to $\rho_0$, making the decomposition similar to $\delta_{n, n'}$. With the further increase of OAM shown in Fig. \ref{fig:Decomp0_25}, $\rho_{\text{L}}$ becomes larger than $\rho_0$, and, once again, higher-order Landau modes appear. In this case, all the Landau states have a larger size than the $\operatorname{NSLG}_{\text{H}}$ state at the boundary. However, their destructive interference results in size suppression (see. Eq. \eqref{DecPhase}).

In Figs. \ref{fig:Decomp0_0_} --- \ref{fig:Decomp12_0} (middle panel, in blue), we set $l = 0$ and investigate how $n$ affects the probability coefficients. In general, the distribution of $\abs{a_{n n' l}}^2$ is similar to that in Figs. \ref{fig:Decomp0_0} --- \ref{fig:Decomp0_25} in the following sense. With increasing $n$, $\rho_{\text{L}}$ grows, and for $n = 7$, when $\rho_{\text{L}} \approx \rho_0$, a $\delta$-like peak emerges in Fig. \ref{fig:Decomp7_0} in accordance with Eq. \eqref{decs}. With a further increase in $n$, this peak vanishes, leading to numerous Landau states in Fig. \ref{fig:Decomp12_0}.

Figs. \ref{fig:Decomp0_35} --- \ref{fig:Decomp3_35} (bottom panel, in green) demonstrate another peculiarity of the probability coefficients distribution. Namely, for sufficiently wide distributions, the number of peaks equals $n + 1$. We suppose this might be connected to the number of rings of the $\operatorname{NSLG}_{\text{H}}$ state; however, the true nature of this phenomenon is still unclear to us.

\section{Emittance}
\label{secEmittance}

\subsection{Emittance and the Schr\"{o}dinger uncertainty relation}

Classical accelerator physics mainly focuses on particle beams, described by distribution functions in phase space. At any moment of time (or any distance $z$ along the direction of beam propagation), every particle in a beam is a point in this space. In systems with axial symmetry, dynamics in two transverse directions are independent and indistinguishable when the beam has no classical vorticity \cite{Groening2021}. This allows monitoring only one transverse coordinate $x(s)$ and the corresponding velocity projection $x'(s)$, which form two-dimensional trace space $(x(s),x'(s))$. Here, $s$ is a variable parametrizing the particle motion, e.g., time or longitudinal coordinate.

\textit{Emittance} is one of the essential measured parameters describing a beam. Depending on the problem, it can be defined in different ways; but the most common definitions are the trace space area and the r.m.s.~emittance \cite{Reiser,PhysRevSTAB.16.011302}. The latter is
\begin{equation}
\label{RMSemit}
    \epsilon_x = \sqrt{\expv{x^2}\expv{x'^2}-\expv{xx'}^2},
\end{equation}
with averaging performed over the beam distribution function,  and $\expv{x} = \expv{x'} = 0$ is assumed. Due to the Liouville's theorem, the phase space volume (or the trace space area) is conserved, but such a definition of emittance does not distinguish between different particle distributions in beams with the same area. Vice versa, the r.m.s.~emittance is not generally constant in time, however, it is sensitive to the beam distribution \cite{Reiser}. One of the reasons why the r.m.s.~emittance depends on time is beam mismatch, which leads to r.m.s.~radius oscillations \cite{Reiser, Noble1989}.

We will now draw analogies between quantum mechanics and classical accelerator physics. While in the latter, particles are points in the phase space, in quantum theory, a single particle packet is smeared in the coordinate and momentum spaces. In quantum mechanics, a quantity similar to that given by Eq.~\eqref{RMSemit} arises from the Schr\"{o}dinger uncertainty relation \cite{Schrodinger1930, Karlovets2021Vortex}
\begin{equation}
    (\Delta \hat{a})^2(\Delta\hat{b})^2 \geq \left(\frac{1}{2}\langle\{\hat{a},\hat{b}\}\rangle-\langle\hat{a}\rangle\langle\hat{b}\rangle\right)^2+\frac{1}{4}\left|\langle[\hat{a},\hat{b}]\rangle\right|^2,
\end{equation}
where $\hat{a}$ and $\hat{b}$ are Hermitian operators. A more illustrative form of this inequality is
\begin{equation}
\label{Schr_ineq}
\begin{aligned}
    (\Delta \hat{a})^2(\Delta\hat{b})^2 - \left(\langle \hat{a}\hat{b} \rangle - \langle\hat{a}\rangle\langle\hat{b}\rangle\right)&\left(\langle \hat{b}\hat{a} \rangle - \langle\hat{b}\rangle\langle\hat{a}\rangle\right) \ge 0.
\end{aligned}
\end{equation}
Note that for $\expv{\hat{a}} = \langle \hat{b}\rangle = 0$, the l.h.s.~of Eq.~\eqref{Schr_ineq} has the same form as the r.h.s.~of Eq.~\eqref{RMSemit}. Thus, when $\hat{a}$ and $\hat{b}$ are the transverse coordinate and velocity operators, respectively, it is natural to call the square root of the l.h.s.~of Eq.~\eqref{Schr_ineq} the \textit{quantum r.m.s. emittance}, see \cite{Karlovets2021Vortex} for more detail. This way, we see that the r.m.s. emittance definition can be naturally extended to quantum mechanics.

In classical physics, the smaller the r.m.s.~emittance is, the less disordered is the beam. In quantum mechanics, the r.m.s.~emittance acquires a new meaning: it reflects \textit{non-classicality} of the state. When the emittance is vanishing, the position-momentum uncertainty is minimal, similar to a classical particle, whose momentum and coordinate can both be measured with minimal error. In contrast, the larger the quantum emittance is, the more noticeable the quantum nature of the particle becomes.

\subsection{Quantum emittance of Laguerre-Gaussian wave~packets}

We now derive the quantum r.m.s.~emittance of the $\operatorname{NSLG}_{\text{f}}$ and $\operatorname{NSLG}_{\text{H}}$ states:
\begin{equation}
\label{Emitfh}
    \epsilon_i = \sqrt{\langle x_i^2\rangle \langle \hat{v}_i^2\rangle - \langle x_i \hat{v}_i\rangle \langle \hat{v}_i x_i \rangle } = \frac{1}{2} \sqrt{\langle \bm{\rho}^2\rangle \langle \hat{\bm{v}}^2\rangle - \langle \bm{\rho} \cdot \hat{\bm{v}}\rangle \langle \hat{\bm{v}} \cdot \bm{\rho} \rangle } \equiv \frac{\epsilon}{2}.
\end{equation}
Here, $i$ enumerates the two transverse axes. The second equality stems from the axial symmetry, and $\hat{\bm{v}} = -i\lambda_{\text{C}}(\nabla - ie \bm{A})$ is the kinetic velocity operator. In Eq.~\eqref{Emitfh}, the averaging is performed over the $\operatorname{NSLG}_{\text{f}}$ or the $\operatorname{NSLG}_{\text{H}}$ states to obtain the corresponding emittance.

\begin{figure}[b]
\subfloat[]{\includegraphics[width=0.4\textwidth]{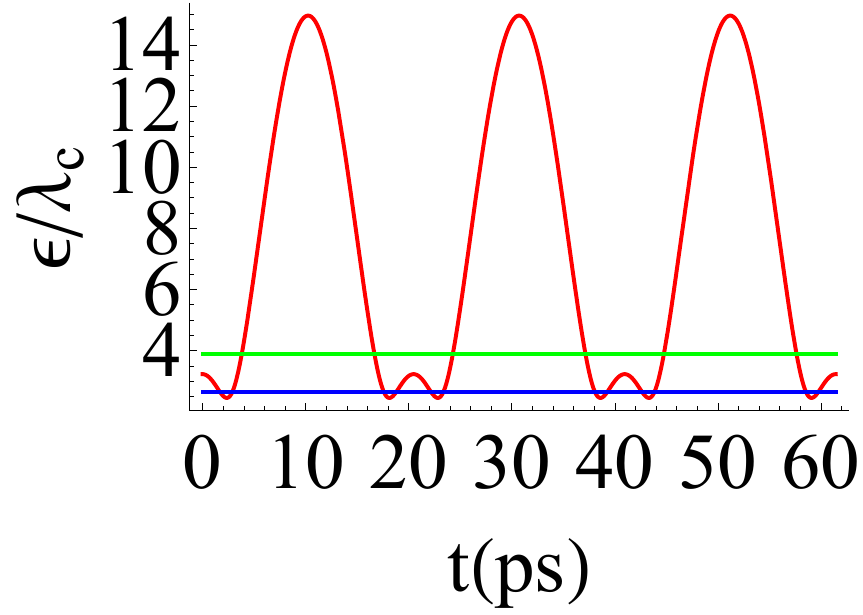} \label{fig:Emitl3}} 
\,
\subfloat[]{\includegraphics[width=0.4\textwidth]{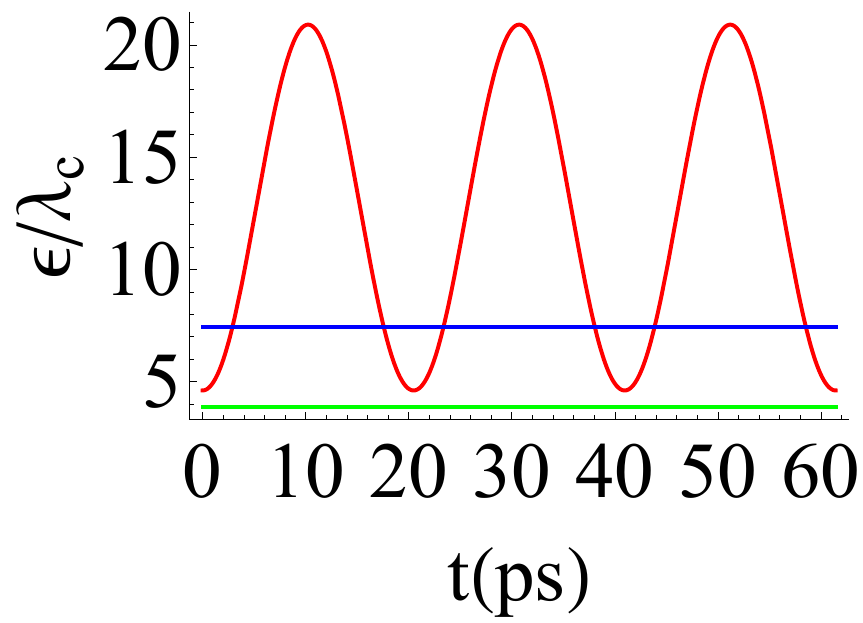} \label{fig:Emitl-3}}
\caption{Emittances of the $\operatorname{NSLG}_{\text{H}}$ (in red), the Landau (in blue), and the $\operatorname{NSLG}_{\text{f}}$ (in green) states. $B = 1.9$ T, $n = 0$, $\sigma_0 = 25$ nm, $\sigma' = 0$. (a) $l = -3$, (b) $l = 3$.}
\label{fig:Emit}
\end{figure}

Using
\begin{equation}
\expv{\bm{\rho}\cdot\hat{\bm{v}}} = \expv{\hat{\bm{v}}\cdot\bm{\rho}}^* = \frac{1}{2}\partial_t\langle\rho^2\rangle (t) + i\lambda_{\text{C}},
\end{equation}
the r.m.s.~emittance can be expressed through the r.m.s.~radius, its derivative, and the average energy as
\begin{equation}
\label{emit}
    \epsilon = \sqrt{2 \lambda_{\text{C}} \langle \rho^2\rangle (t)\expv{E}-\frac{1}{4}\left[\partial_t \expv{\rho^2}(t)\right]^2 -\lambda_{\text{C}}^2}.
\end{equation}
Let us first focus on the $\operatorname{NSLG}_{\text{f}}$ state. By substituting explicit expressions for the wave packet parameters from Eqs. \eqref{r.m.s.RadFree} and \eqref{FLGEnEx} into Eq.~\eqref{emit}, we get
\begin{equation}
\label{emf}
    \epsilon_{\text{f}} = \lambda_{\text{C}}\sqrt{(2n+|l|+1)^2 - 1}.
\end{equation}
The r.m.s.~emittance of a free particle is constant in time and minimal for the Gaussian electron state when $n = l = 0 $. Notice that this state minimizes the Schr\"{o}dinger uncertainty, not the Heisenberg one. We should note this state is a special case of the coherent states of a free particle discussed in \cite{Bagrov2014}. For $n$, $|l| \sim 1$, the quantum emittance of an $\operatorname{NSLG}_{\text{f}}$ state is of the order of $\lambda_{\text{C}}$, i.e. the particle stays relatively ``classical''. For large quantum numbers, the emittance grows linearly, and the quantum nature of the particle becomes more pronounced. 

Similarly, using $\operatorname{NSLG}_{\text{H}}$ optical functions and energy discussed in Sec. \ref{E}, we obtain the r.m.s.~emittance of an NSLG electron inside the solenoid:
\begin{equation}
\label{emt}
     \epsilon_{{\text{H}}} (t) = \lambda_{\text{C}} \sqrt{\frac{\epsilon_{\text{f}}^2}{\lambda_{\text{C}}^2} + \left[ (2n + \abs{l} + 1) \frac{\sigma^2(t)}{\sigma_{\text{L}}^2} + l \right]^2 - l^2}.
\end{equation}
The r.m.s.~emittance of an $\operatorname{NSLG}_{\text{H}}$ state is defined by the dispersion $\sigma(t)$. The time dependence stems from the mismatch at the boundary ($\sigma_0 \ne \sigma_{\text{L}}$ and/or $\sigma'_0 \neq 0$), which causes the r.m.s.~radius and, hence, the r.m.s.~emittance oscillations.

From Eq.~\eqref{emt}, the r.m.s.~emittance of the Landau state can be easily obtained by setting $\sigma(t) = \sigma_{\text{L}}$:
\begin{equation}
\label{emtL}
    \epsilon_{{\text{L}}} = \lambda_{\text{C}} \sqrt{\frac{\epsilon_{\text{f}}^2}{\lambda_{\text{C}}^2} + (2n + \abs{l} + l + 1)^2 - l^2}.
\end{equation}
One can notice that the r.m.s.~emittance is discontinuous at the boundary. This can be seen from Eq.~\eqref{emit}: the dispersion and its derivative are continuous, while the average energy is not, as we discussed in the beginning of Sec. \ref{secNSLGandFree}.

The time dependence of the $\operatorname{NSLG}_{\text{H}}$ emittance is shown in Fig. \ref{fig:Emit}. Unlike the r.m.s.~radius, it is sensitive to the OAM sign. For $l < 0$ (Fig. \ref{fig:Emitl3}), r.m.s.~emittance has additional local maxima, in contrast to the case when the OAM and the magnetic field are aligned (Fig. \ref{fig:Emitl-3}).
 
Following the idea that smaller quantum r.m.s.~emittance corresponds to a ``more classical'' particle behavior, we will analyze the regime when $\epsilon_{{\text{H}}}(t) < \epsilon_{{\text{f}}}$. For $n$, $|l| \sim 1$, it means that $\epsilon_{\text{H}} \lesssim 1$. Fig.
\ref{fig:Emit} shows that for some parameters of the wave packet, there are time intervals when this condition is satisfied. From Eq.~\eqref{emt}, this is possible only for $l < 0$. Moreover, the following relation has to be fulfilled:
\begin{equation}
    \frac{2n + \abs{l} + 1}{4 \abs{l}} + \frac{\abs{l}}{2n + \abs{l} + 1}  < \frac{\sigma_{\text{st}}^2}{\sigma^2_{\text{L}}} < \frac{2 \abs{l}}{2n + \abs{l} + 1}.
\end{equation}
Note that for $n$ or $l \gg 1$, $\operatorname{NSLG}_{\text{H}}$ emittance greatly exceeds $\lambda_{\text{C}}$ when these inequalities are violated.

Therefore, the emittance of an $\operatorname{NSLG}$ electron can be locally decreased  if the electron is placed in the field. However, if we consider a finite-length solenoid, the emittance changes abruptly at both boundaries, and when the particle leaves the solenoid, the emittance is exactly the same as it was at the entrance. Thus, our findings open ways for altering the r.m.s.~emittance of an electron with magnetic lenses.

\section{Results and discussion}
\label{secConclusion}

We have analyzed the properties of nonstationary Laguerre-Gaussian (NSLG) states, which, unlike the Landau states, fully capture vortex electron dynamics both at the vacuum-solenoid boundary and inside the magnetic field. Wave functions of an electron in free space and in the magnetic field belong to the same class of functions, which enables a smooth transition between single-mode states with the same quantum numbers.

The vector potential of the magnetic field was chosen in the symmetric gauge, which has led us to the Laguerre-Gaussian states. However, an alternative choice of the vector potential gauge would result in a different family of states, such as Hermite-Gaussian states. Which gauge to use is determined by the initial state of an electron in free space and, therefore, by the boundary conditions.

The decomposition of the NSLG states in a solenoid into the conventional basis of the Landau states was performed. A wave packet slightly mismatched with a Landau state at the boundary propagates through the magnetic lens as a superposition of a few Landau states with \textit{the same} OAM and neighbouring radial quantum numbers. In other cases the electron further propagates in the field as a complex superposition of Landau states with the OAM of the initial state but significantly different radial quantum numbers.

We have considered a twisted electron entering the solenoid at a small angle $\alpha$ to the field direction. For any sensible values of the electron energy and momentum, the condition $\alpha \ll 1$~rad is sufficient to neglect any corrections to a single NSLG state in a solenoid. Thus, the OAM of the quantum packet is robust against little deviations from the axial symmetry and small inhomogeneties of the field, which supports our previous findings \cite{Karlovets2021Vortex}.

Our calculations show that transversely relativistic and longitudinally nonrelativistic beams of twisted particles can be achieved in existing experimental setups. For instance, electrons with quantum numbers of the order of $10^3$, generated as NSLG states with a waist size of 1~$\mu$m and focused afterwards to 1 nm, become transversely relativistic. Such particles can be a curious object of study in accelerator physics, as their dynamics significantly differs from that of regular accelerator beams. Moreover, the significance of these transversely relativistic states cannot be overstated when it comes to investigating radiation and self-polarization phenomena in synchrotrons. Nonetheless, such application of NSLG states necessitates their relativistic extension.

Finally, we have introduced the quantum analogue of beam emittance for a quantum wave packet and applied it to the NSLG state. This quantity explicitly measures the non-classicality of the state via the Schr\"{o}dinger uncertainty relation, which is more general than the well-known Heisenberg inequality. The quantum emittance of an NSLG state grows linearly with $n$ and $l$ for large quantum numbers. In free space, for fundamental Gaussian mode ($n = l = 0$), the emittance vanishes, or, equivalently, the Schr\"{o}dinger uncertainty relation turns into equality. This reflects the semiclassical character of the Gaussian state and the ``quantumness'' of the wave packets with large quantum numbers $n$ and $l$. For an electron inside the field, the emittance generally oscillates in time, and for negative OAM, it can be locally lower than the emittance of a free NSLG state that enters the lens.

\section{Conclusion}

Let us summarize the main results. The Landau states serve as a convenient tool for describing electron motion in condensed matter or photon emission in magnetic fields. However, once particles can transfer between vacuum and magnetic field regions NSLG states appear as a more advantageous means for describing their states. NSLG states allow to account for such a transition of a vortex electron in terms of a single mode --- a state with given quantum numbers, the same to the both sides of the boundary. Moreover, the electron dynamics can be straightforwardly extracted from the NSLG state itself, whereas using the Landau states requires complex superposition of them and further calculations of observables.

The nonstationary nature of the transition process is imprinted in the time dependence of the NSLG wavefunctions, leading to oscillations in the electron's r.m.s. radius entering a field. The oscillation parameters can be set via boundary conditions controlled by an experimenter. For example, to obtain an electron in a Landau state, the r.m.s. radius of the incoming electron state at the boundary must be adjusted to that of the Landau state with the same quantum numbers, and the divergence rate must be vanishing. Along with that, the time dependence of the r.m.s. radius of a wave packet can be utilized to locally enhance focusing and is of great importance for precise measurements.

While NSLG states have already demonstrated their advantages for electron microscopes and linacs, we see two key directions for elaborating upon this model. First, generalizing this model to higher energies is necessary for applying the NSLG states approach to the dynamics of vortex electrons with arbitrarily large angular momentum, where the transverse dynamics becomes relativistic. Second, a more realistic description involves accounting for the finite lifetime of NSLG states due to photon emission. In addition to spontaneous radiative transitions between states with different quantum numbers present for both the Landau and NSLG states. The nonstationary nature of the latter introduces another radiative mechanism --- the classical quadrupole radiation caused by the oscillations in the r.m.s. radius. A comprehensive exploration of these aspects could extend the NSLG states approach to a variety of applications as, for example, generating vortex gamma photons using twisted electrons in accelerators.

\section*{Acknowledgement}

We are grateful to S.\,Baturin, A.\,Volotka, and D.\,Glazov for the fruitful discussions and criticism. The studies in Secs. II are supported
by the Russian Science Foundation (Project No. 21-42-04412;
https://rscf.ru/en/project/21-42-04412/). The studies in Sect. III are supported by the Ministry of Science and Higher Education of the Russian Federation (agreement No.075-15-2021-1349). The work on the quantum states (by D. Karlovets, G. Sizykh, and D. Grosman) in Sec.IV was supported by the Foundation for the Advancement of Theoretical Physics and Mathematics “BASIS”. The studies in Sec. V are supported by the Government of the Russian Federation through the ITMO Fellowship and Professorship Program. The studies in Sec.VI are supported by the Russian Science Foundation (Project No.\,23-62-10026;  https://rscf.ru/en/project/23-62-10026/).

\appendix

\section{Completeness of nonstationary Laguerre-Gaussian states}
\label{sec:Completeness}

We can prove that the set of states \eqref{NSLG} is complete the following way. Let us consider a moment of time $t = \mathcal{T}$ such that $\sigma(\mathcal{T}) = \sigma$ and $\sigma'(\mathcal{T}) = 0$. This corresponds to $R(\mathcal{T}) \rightarrow \infty$, and the wave function \eqref{NSLG} takes the following form:
\begin{equation}
\label{co}
    \Psi_{n l}(\bm{\rho}, \mathcal{T}) = \Psi^{(\text{L})}(\rho,\varphi,t)\exp(-i\Phi_G(\mathcal{T}) + iE_{\text{L}}t).
\end{equation}
Here, $\Psi^{(\text{L})}(\rho,\varphi,t)$ are the wave functions of the Landau states in an effective magnetic field $H_{\text{eff}} = 2/(e_0 \sigma^2)$, which are complete in $\mathcal{L}^2(\mathbb{R})$, and $t$ is an arbitrary moment of time. Completeness of Hermite-Gaussian functions is proven in Theorem 11.4 in \cite{Hall}, which can be directly adopted to the Laguerre-Gaussian counterparts $\Psi^{(\text{L})}(\rho,\varphi,t)$. From Eq. \eqref{co}, it is clear that if $c_{n l}(t)$ are the coefficients of decomposition of some function of $(\bm{\rho},t)$ by Landau states, then $\tilde{c}_{n l}(t) = c_{n l}(t)\exp(i\Phi_G(\mathcal{T}) - iE_{\text{L}}t)$ are the decomposition coefficients for the same function into NSLG states \eqref{NSLG} evaluated at a time $t = \mathcal{T}$. Let us now decompose some function $F(\bm{\rho},t)$ into the wave functions \eqref{NSLG} at an \textit{arbitrary} moment of time. First, we consider another function $G(\bm{\rho},t) = \exp(i\hat{\mathcal{H}}(t-\mathcal{T}))F(\bm{\rho},t)$, where $\hat{\mathcal{H}}$ is the transverse part of the Hamiltonian of an electron in free space or in a magnetic field. The function $G(\bm{\rho},t)$ can be uniquely decomposed into Landau states and, hence, into NSLG states evaluated at a time $t = \mathcal{T}$:
\begin{equation}
\label{co1}
    G(\bm{\rho},t) = \exp(i\hat{\mathcal{H}}(t-\mathcal{T}))F(\bm{\rho},t) = 
    \sum\limits_{n,l}c_{n l}(t)\Psi^{\text{(L)}}(\rho,\varphi,t) = 
    \sum\limits_{n,l}\tilde{c}_{n l}(t)\Psi_{n l}(\bm{\rho},\mathcal{T}).
\end{equation}
Acting on both sides of Eq.~\eqref{co1} with $\exp(-i\hat{\mathcal{H}}(t-\mathcal{T}))$, which is the evolution operator for the states \eqref{NSLG}, we obtain\begin{equation}
    F(\bm{\rho},t) = \sum\limits_{n l}\tilde{c}_{n l}(t)\Psi_{n l}(\bm{\rho},t),
\end{equation}
which proves the completeness of the NSLG states.

Moreover, if the functions to be decomposed and the NSLG states satisfy the Schr\"{o}dinger equation with the same Hamiltonian, and, thus, their time dependence is governed by the same evolution operator, the decomposition coefficients are independent of time. Indeed, consider the decomposition
\begin{equation}
\label{notime}
    \Psi(\bm{\rho},t) = \sum\limits_{n,l}c_{n l}(t)\Psi_{n l}(\bm{\rho},t),
\end{equation}
where $\Psi(\bm{\rho},t)$ and $\Psi_{n l}(\bm{\rho},t)$ satisfy the same Schr\"{o}dinger equation. Since Eq. \eqref{notime} is valid for any moment of time, we also have the following decomposition:
\begin{equation}
\label{notime1}
    \Psi(\bm{\rho},0) = \sum\limits_{n,l} c_{n l}(0)\Psi_{n l}(\bm{\rho},0).
\end{equation}

Acting on both sides of Eq.~\eqref{notime1} with the evolution operator, which does not affect the decomposition coefficients, we arrive at
\begin{equation}
    \Psi(\bm{\rho},t) = \sum\limits_{n,l}c_{n l}(0)\Psi_{n l}(\bm{\rho},t).
\end{equation}
Now we recall that the set of the NSLG states is complete, and the choice of the decomposition coefficients is unique, meaning $c_{n,l}(t) = c_{n,l}(0)$, which, in turn, implies that the coefficients in this case do not depend on time.

\section{Influence of divergence rate on oscillations}
\label{DivRateInf}

\begin{figure*}[b]
 \begin{subfigure}{0.24\textwidth}
     \includegraphics[width=\textwidth]{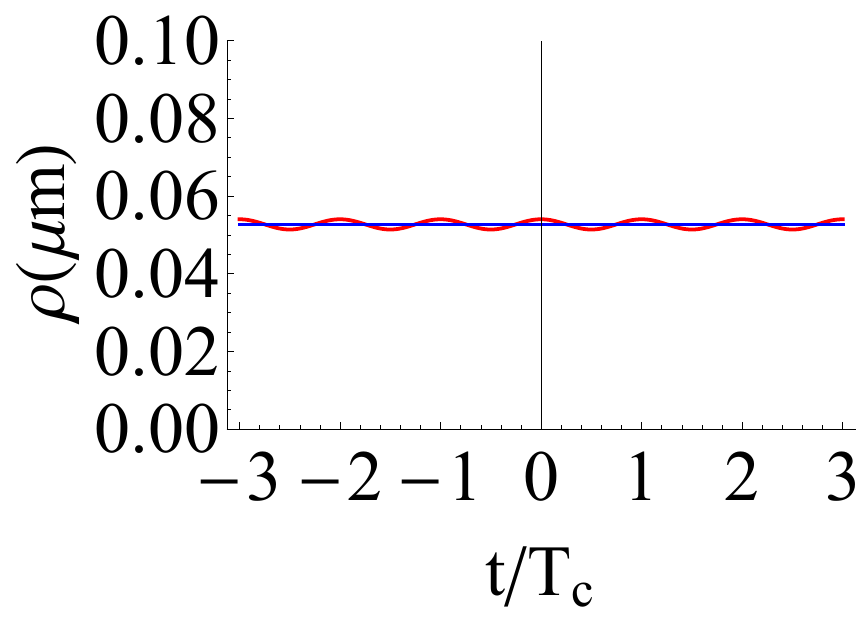}
     \caption{}
     \label{fig:NSLG_BSZero}
 \end{subfigure}
 \hfill
 \begin{subfigure}{0.24\textwidth}
     \includegraphics[width=\textwidth]{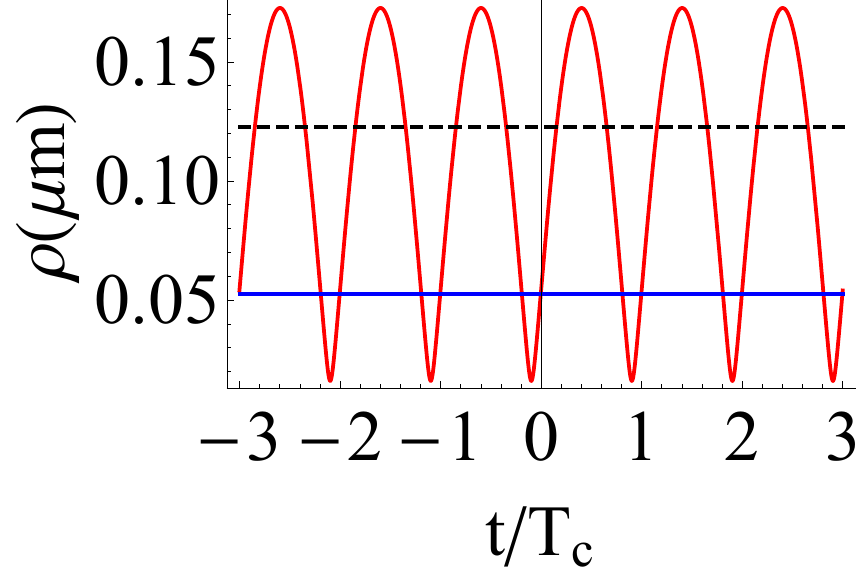}
     \caption{}
     \label{fig:NSLG_BSLow}
 \end{subfigure}
 \hfill
 \begin{subfigure}{0.24\textwidth}
     \includegraphics[width=\textwidth]{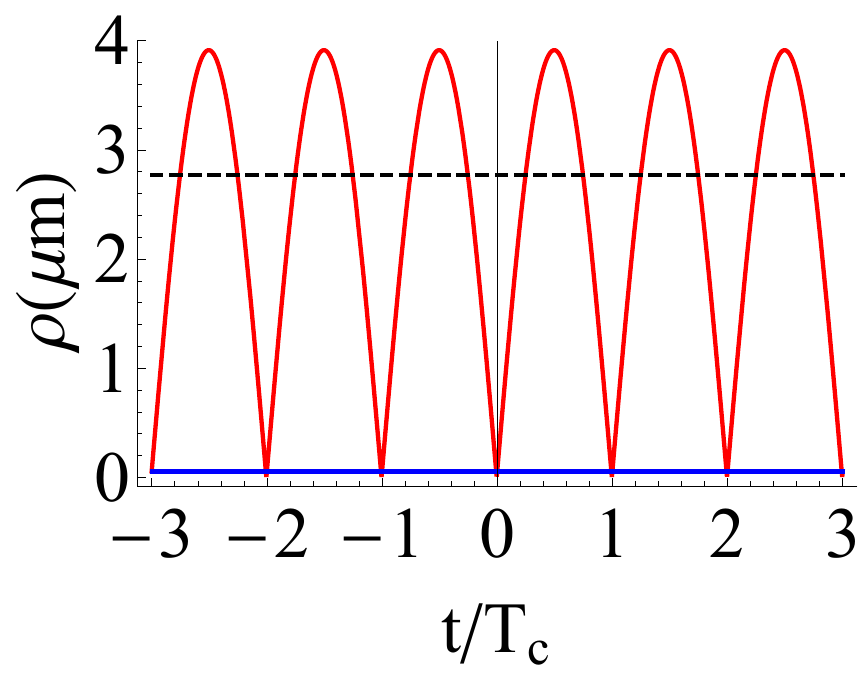}
     \caption{}
     \label{fig:NSLG_BSMed}
 \end{subfigure}
 \hfill
 \begin{subfigure}{0.24\textwidth}
     \includegraphics[width=\textwidth]{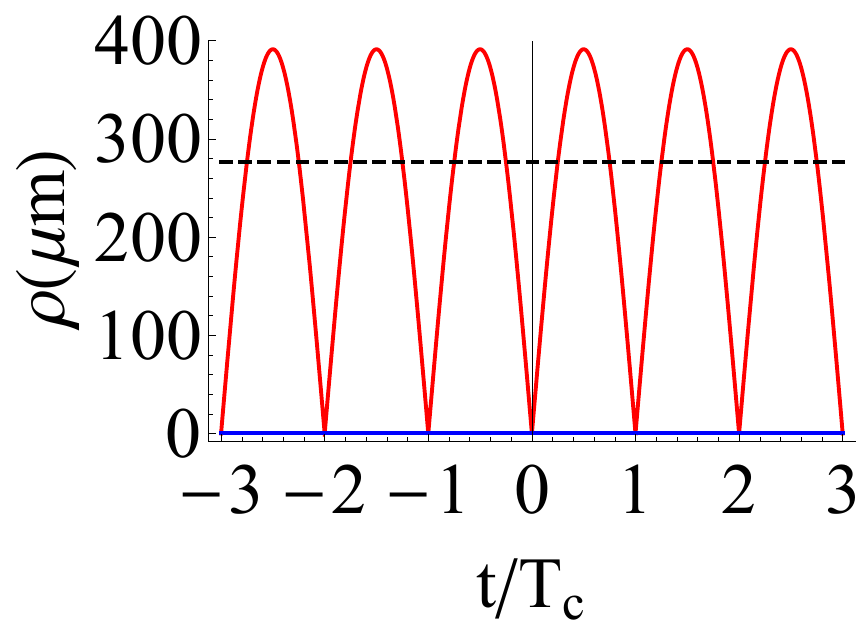}
     \caption{}
     \label{fig:NSLG_BSHigh}
 \end{subfigure}

\caption{The $\operatorname{NSLG}_{\text{H}}$ packet r.m.s. radii (in red) for different $\rho'_0$. The Landau radius $\rho_{\text{L}}$ is in blue. Black dashed lines correspond to $\rho_{\text{st}}$ given by Eq. \eqref{StR}. In each subfigure $H = 1.9$ T ($T_{\text{c}} \approx 0.02$ ns), n = 0, l = 3, $\rho_0 = 54$ nm. (a) $\rho'_0 = 0$, (b) $\rho'_0 = 4 \times 10^{-5}$, (c) $\rho'_0 = 10^{-3}$, (d) $\rho'_0 = 0.1$.}
 \label{fig:DivergenceRateInfluence}
\end{figure*}

The initial divergence rate of the $\operatorname{NSLG}_{\text{H}}$ wave packet significantly influences its r.m.s.~radius oscillations. The effect is depicted in Fig. \ref{fig:DivergenceRateInfluence}. We choose the parameters as follows: $H = 1.9$ T, $n = 0$, $l = 3$, $\rho_0 = 25$ nm.

Fig. \ref{fig:NSLG_BSZero} serves as a reference with $\rho'_0 = 0$. In Fig. \ref{fig:NSLG_BSLow}, the divergence rate $\rho'_0 = 4 \times 10^{-5}$ such that the second and the third terms of $\sigma_{\text{st}}^2$ in Eq. \eqref{NSLGDisp} both contribute to its value. The nonzero divergence rate leads to a little shift in the initial phase of the oscillations $\theta$ and increase of their amplitude. Note that a change in the sign of $\rho'_0$ does not alter the amplitude and simply results in the phase shift with the opposite sign according to Eqs. \eqref{NSLGDisp}. In Fig. \ref{fig:NSLG_BSMed}, the divergence rate is $\rho'_0 = 10^{-3}$. For such a high value of $\rho'_0$, the initial phase of oscillations is negligible, and the amplitude is enhanced even more. In this regime, the magnitude grows proportionally to $\rho'_0$, that is clear from the comparison of Figs. \ref{fig:NSLG_BSMed} and \ref{fig:NSLG_BSHigh}.

\section{Vanishing magnetic field limit of optical functions}
\label{Limit}
Consider the dispersion of the $\operatorname{NSLG}_{\text{H}}$ state. Let us rewrite it by substituting the phase of oscillations $\theta$:
\begin{equation}
\label{limit}
\begin{aligned}
    \sigma^2(t) = \sigma_{\text{st}}^2 + \sigma_{\text{st}}^2 & \sqrt{1-\frac{\sigma_{\text{L}}^4}{\sigma_{\text{st}}^4}}\sin(s(\sigma_0,\sigma_0')\omega\tau-\theta) = \sigma_{\text{st}}^2 - \left(\sigma_{\text{st}}^2-\sigma_0^2\right)\cos(\omega\tau)  \\
    &+s(\sigma_0,\sigma_0')  \sin(\omega\tau) \sqrt{2\sigma_0^2\sigma_{\text{st}}^2 + \sigma_0^4 - \sigma_{\text{L}}^4},
\end{aligned}
\end{equation}
where $\tau = t - t_0$. In the vanishing magnetic field limit, when $H \rightarrow 0$ and $\sigma_{\text{L}} \rightarrow \infty$, the stationary dispersion can be simplified to
\begin{equation}
    \sigma_{\text{st}} = \sigma_{\text{L}}^2\left(\frac{1}{2 \sigma_0^2} + \frac{\sigma_0'^2}{2 \lambda_{\text{C}}^2}\right)^{1/2}.
\end{equation}

Now we only keep the nonvanishing terms in Eq. \eqref{limit} in the limit $H \rightarrow 0$:
\begin{equation}
\label{limit1}
    \sigma^2(t) \rightarrow \sigma_0^2 + 2\lambda_{\text{C}}^2\tau^2\left(\frac{1}{2\sigma_0^2} + \frac{\sigma_0'^2}{2 \lambda_{\text{C}}^2}\right) + \frac{2 \sigma_0'\tau}{\sigma_0}.
\end{equation}
Then we express $\sigma_0$ and $\sigma_0'$ via the waist dispersion and the diffraction time as
\begin{equation}
\label{limit2}
\sigma_0 = \sigma_{\text{w}}\sqrt{1+\frac{(t_0 - t_{\text{g}})^2}{\tau_{\text{d}}^2}}, \hspace{30pt} \sigma_0' = \frac{\sigma_{\text{w}}^2(t_0 - t_{\text{g}})}{\tau_d^2\sqrt{1+\frac{(t_0 - t_{\text{g}})^2}{\tau_{\text{d}}^2}}}.
\end{equation}
Substituting Eq. \eqref{limit2} into Eq. \eqref{limit1} yields
\begin{equation}
    \sigma(t) = \sigma_{\text{w}}\sqrt{1+\frac{(t_0 - t_{\text{g}})^2}{\tau_{\text{d}}^2}}.
\end{equation}

Smooth transformations of other optical functions follow from the system of optical equations \eqref{eq:Opt} and transformation of the dispersion demonstrated above.

\section{Explicit form of decomposition coefficients}

The coefficients of the $\operatorname{NSLG}_{\text{H}}$ state decomposition into stationary Landau wave functions are given by the following integral:
\begin{equation}
\begin{aligned}
    & a_{n n' l} = |N_{nl}|^2 \int\limits_{0}^{\infty} \frac{\rho^{2|l|}}{(\sigma_{\text{L}}\sigma(t))^{|l|+1}}L_{n'}^{|l'|}\left[\frac{\rho^2}{\sigma_{\text{L}}^2}\right]L_{n}^{|l|}\left[\frac{\rho^2}{\sigma^2(t)}\right] \\
    \times \exp & \biggl[ - \left(\frac{\rho^2}{2 \sigma_{\text{L}}^2} + \frac{\rho^2}{2 \sigma^2(t)}\right) + i\frac{\rho^2}{2\lambda_{\text{C}}R(t)} - i\Phi_{\text{G}}(t) + i E_{\text{L}}(t - t_0) \biggr]  d^2\rho.
\end{aligned}
\end{equation}
The coefficients are independent of time and can be evaluated at $t = t_0$ for simplicity, when $\sigma(t_0) = \sigma_0, R(t_0) = \sigma_0 / \sigma_0'$ and $\Phi_{\text{G}}(t_0) = \Phi_0$. 
\label{NSLGDecLandauApp}

The integral can be evaluated using Eq.(7.422) in \cite{Gradshtein} (there is, however, a misprint $m\leftrightarrow n$) and presented in the following form:
\begin{equation}
\label{Dec}
    a_{n n' l} = (\zeta^2-1)^{(n'-n)/2}g(\zeta)e^{i\chi_{n n'}}.
\end{equation}
Here $\zeta = \rho_{\text{st}}/\rho_{\text{L}}$,
\begin{equation}
\begin{aligned}
    g(\zeta) & = \frac{(n+n'+|l|)!}{\sqrt{n!n'!(n+|l|)!(n'+|l|)!}}\frac{(-2)^{n}}{(\lambda+1)^{(n+n'+|l|+1)/2}} \\
    & \times \left| \tensor[_2]{F}{_1}\left[ -n, -n-|l|; -n - n' - |l|; \frac{\zeta^2}{2}+\frac{1}{2}\right]\right|
\end{aligned}
\end{equation}
is an analytic function, and the phase
\begin{equation}
\label{DecPhase}
\begin{aligned}
    & \chi_{n n' l} = 
        \Phi_0 + \left\{
        \begin{array}{ll}
        0, & \mbox{if $\tensor[_2]{F}{_1} \geq 0$}.\\
        \pi, & \mbox{if $\tensor[_2]{F}{_1} < 0$}.
        \end{array}
        \right\} + 
        \pi \times
        \left\{
        \begin{array}{ll}
        n, & \mbox{if $\xi_1 < 1$}\\
        n', & \mbox{if $\xi_1 > 1$}
        \end{array}
        \right\} \\
        & + (n - n') \arctan{\frac{\xi_1 \xi_2}{1-\xi_1^2}} + 
    (n + n' + |l| + 1) \arctan{\frac{\xi_1 \xi_2}{1+\xi_1^2}}.
\end{aligned}
\end{equation}
In the limit $\zeta^2 \rightarrow 1$
\begin{equation}
\begin{cases}
    g(\zeta) \propto 1, \;\text{if} \; n' > n, \\
    g(\zeta) \propto (\zeta^2 - 1)^{n-n'}, \;\text{if} \; n > n',
\end{cases}
\end{equation}
which provides the following asymptotic for the decomposition coefficients:
\begin{equation}
    a_{n n' l} \propto (\delta \zeta)^{\frac{|n'-n|}{2}}.
\end{equation}

\bibliographystyle{unsrt}
\bibliography{references}

\begin{thebibliography}{10}

\bibitem{Bliokh2007}
Konstantin~Yu. Bliokh, Yury~P. Bliokh, Sergey Savel'ev, and Franco Nori.
\newblock Semiclassical dynamics of electron wave packet states with phase
  vortices.
\newblock {\em Phys. Rev. Lett.}, 99:190404, Nov 2007.

\bibitem{Bliokh2011}
Konstantin~Y. Bliokh, Mark Dennis, and Franco Nori.
\newblock Relativistic electron vortex beams: Angular momentum and spin-orbit
  interaction.
\newblock {\em Phys. Rev. Lett.}, 107:174802, Oct 2011.

\bibitem{Bliokh2012}
Konstantin~Y. Bliokh, Peter Schattschneider, Jo~Verbeeck, and Franco Nori.
\newblock {Electron Vortex Beams in a Magnetic Field: A New Twist on Landau
  Levels and Aharonov-Bohm States}.
\newblock {\em Phys. Rev. X}, 2:041011, Nov 2012.

\bibitem{Gallatin2012}
Gregg~M. Gallatin and Ben McMorran.
\newblock Propagation of vortex electron wave functions in a magnetic field.
\newblock {\em Phys. Rev. A}, 86:012701, Jul 2012.

\bibitem{Karlovets2012}
Dmitry Karlovets.
\newblock Electron with orbital angular momentum in a strong laser wave.
\newblock {\em Phys. Rev. A}, 86:062102, Dec 2012.

\bibitem{Greenshields2012}
Colin Greenshields, Robert~L Stamps, and Sonja Franke-Arnold.
\newblock Vacuum {Faraday} effect for electrons.
\newblock {\em New Journal of Physics}, 14(10):103040, Oct 2012.

\bibitem{Ivanov2016}
I.~Ivanov, V.~Serbo, and V.~Zaytsev.
\newblock Quantum calculation of the vavilov-cherenkov radiation by twisted
  electrons.
\newblock {\em Phys. Rev. A}, 93:053825, May 2016.

\bibitem{Karlovets2018}
Dmitry Karlovets.
\newblock Relativistic vortex electrons: Paraxial versus nonparaxial regimes.
\newblock {\em Phys. Rev. A}, 98:012137, Jul 2018.

\bibitem{Maiorova2018}
A.~Maiorova, S.~Fritzsche, R.~Müller, and A.~Surzhykov.
\newblock Elastic scattering of twisted electrons by diatomic molecules.
\newblock {\em Phys. Rev. A}, 98:042701, Oct 2018.

\bibitem{Ducharme2021}
Robert Ducharme, Irismar da~Paz, and Armen Hayrapetyan.
\newblock {Fractional Angular Momenta, Gouy and Berry Phases in Relativistic
  Bateman-Hillion-Gaussian Beams of Electrons}.
\newblock {\em Phys. Rev. Lett.}, 126:134803, Apr 2021.

\bibitem{Karlovets2021Nonlinear}
D.~Karlovets and A.~Pupasov-Maksimov.
\newblock Nonlinear quantum effects in electromagnetic radiation of a vortex
  electron.
\newblock {\em Phys. Rev. A}, 103:012214, Jan 2021.

\bibitem{Shaohu2021}
Lei Shaohu, Bu~Zhigang, Wang Weiqing, Shen Baifei, and Ji~Liangliang.
\newblock Generation of relativistic positrons carrying intrinsic orbital
  angular momentum.
\newblock {\em Phys. Rev. D}, 104:076025, Oct 2021.

\bibitem{Verbeeck2010}
J.~Verbeeck, H.~Tian, and Peter Schattschneider.
\newblock Production and application of electron vortex beams.
\newblock {\em Nature Lett.}, 467:301--304, Sep 2010.

\bibitem{McMorran2011}
Benjamin~J. McMorran, Amit Agrawal, Ian~M. Anderson, Andrew~A. Herzing,
  Henri~J. Lezec, Jabez~J. McClelland, and John Unguris.
\newblock Electron vortex beams with high quanta of orbital angular momentum.
\newblock {\em Science}, 331(6014):192--195, 2011.

\bibitem{Guzzinati2013}
Giulio Guzzinati, Peter Schattschneider, Konstantin Bliokh, Franco Nori, and
  Jo~Verbeeck.
\newblock Observation of the larmor and gouy rotations with electron vortex
  beams.
\newblock {\em Phys. Rev. Lett.}, 110:093601, Mar 2013.

\bibitem{Petersen2013}
T.~Petersen, D.~Paganin, M.~Weyland, T.~Simula, S.~Eastwood, and M.~Morgan.
\newblock Measurement of the gouy phase anomaly for electron waves.
\newblock {\em Phys. Rev. A}, 88:043803, May 2013.

\bibitem{Schattschneider2014}
P.~Schattschneider, Th. Schachinger, M.~St\"{o}ger-Pollach, S.~L\"{o}ffler,
  Steiger-Thirsfeld A., Bliokh K., and F.~Nori.
\newblock Imaging the dynamics of free-electron {Landau }states.
\newblock {\em Nature Comm.}, 5:4586, Aug 2014.

\bibitem{Schachinger2015}
T.~Schachinger, S.~Löffler, Stöger-Pollach M., and P.~Schattschneider.
\newblock Peculiar rotation of electron vortex beams.
\newblock {\em Ultramicroscopy}, 158:17--25, Nov 2015.

\bibitem{Idrobo2011}
J.C. Idrobo and S.J. Pennycook.
\newblock Vortex beams for atomic resolution dichroism.
\newblock {\em Microscopy}, 60:295--300, Oct 2011.

\bibitem{Mohammadi2012}
Z.~Mohammadi, C.P. Van~Vlack, S.~Hughes, J.~Bornemann, and R.~Gordon.
\newblock Vortex electron energy loss spectroscopy for near-field mapping of
  magnetic plasmons.
\newblock {\em Optics Express}, 20:15024--15034, 2012.

\bibitem{Grillo2017}
Vincenzo Grillo, Tyler Harvey, Federico Venturi, Jordan Pierce, Roberto
  Balboni, Frédéric Bouchard, Gian Gazzadi, Stefano Frabboni, Amir Tavabi,
  Zi-An Li, Rafal Dunin-Borkowski, Robert Boyd, Benjamin McMorran, and Ebrahim
  Karimi.
\newblock Observation of nanoscale magnetic fields using twisted electron
  beams.
\newblock {\em Nature Comm.}, 8:689, Sep 2017.

\bibitem{Bliokh2017}
K.Y. Bliokh, I.P. Ivanov, G.~Guzzinati, L.~Clark, R.~{Van Boxem},
  A.~B{\'e}ch{\'e}, R.~Juchtmans, M.A. Alonso, P.~Schattschneider, F.~Nori, and
  J.~Verbeeck.
\newblock Theory and applications of free-electron vortex states.
\newblock {\em Physics Reports}, 690:1--70, 2017.
\newblock Theory and applications of free-electron vortex states.

\bibitem{Karlovets2021Vortex}
Dmitry Karlovets.
\newblock Vortex particles in axially symmetric fields and applications of the
  quantum {Busch} theorem.
\newblock {\em New Journal of Physics}, 23(3):033048, mar 2021.

\bibitem{Greenshields2014}
C.R. Greenshields, R.L. Stamps, S.~Franke-Arnold, and S.M. Barnett.
\newblock Is the angular momentum of an electron conserved in a uniform
  magnetic field?
\newblock {\em Phys. Rev. Lett.}, 113:240404, Dec 2014.

\bibitem{Greenshields2015}
C.R. Greenshields, S.~Franke-Arnold, and R.L. Stamps.
\newblock Parallel axis theorem for free-space electron wavefunctions.
\newblock {\em New J. Phys.}, 17:093015, 2015.

\bibitem{STeng}
A.A. Sokolov and I.M. Ternov.
\newblock {\em Radiation from Relativistic Electrons}.
\newblock American Inst. of Physics, 1986.

\bibitem{BagrovBlackBook}
V.G. Bagrov, G.S. Bisnovatiy-Kogan, V.A. Borodovitsyn, A.V. Borisov, O.F.
  Dorofeev, V.Ch. Zhukovskiy, Yu.L. Pivovarov, O.V. Shorokhov, and V.Ya. Epp.
\newblock {\em Theory of Radiation of Relativistic Particles}.
\newblock Fiz. Mat. Lit., Moscow, 2002.

\bibitem{Landau}
L.~D. Landau and E.~M. Lifshitz.
\newblock {\em Quantum Mechanics: Nonrelativistic Theory}.
\newblock Butterworth-Heinemann, Burlington, Massachusetts, 1981.

\bibitem{Silenko2021}
Zou Liping, Zhang Pengming, and Alexander Silenko.
\newblock General quantum-mechanical solution for twisted electrons in a
  uniform magnetic field.
\newblock {\em Phys. Rev. A}, 103:L010201, Jan 2021.

\bibitem{Melkani2021}
Abhijeet Melkani and S.~J. van Enk.
\newblock Electron vortex beams in nonuniform magnetic fields.
\newblock {\em Phys. Rev. Research}, 3:033060, Jul 2021.

\bibitem{Baturin2022}
S.~Baturin, D.~Grosman, G.~Sizykh, and D.~Karlovets.
\newblock Evolution of an accelerated charged vortex particle in an
  inhomogeneous magnetic lens.
\newblock {\em Phys. Rev. A}, 106:042211, Oct 2022.

\bibitem{Sizykh2023}
G.~K. Sizykh, A.~D. Chaikovskaia, D.~V. Grosman, I.~I. Pavlov, and D.~V.
  Karlovets.
\newblock Transmission of vortex electrons through a solenoid.
\newblock {\em arXiv:2306.13161 [quant-ph]}, June 2023.

\bibitem{Karlovets2019Intrinsic}
Dmitry Karlovets and Alexey Zhevlakov.
\newblock Intrinsic multipole moments of non-gaussian wave packets.
\newblock {\em Phys. Rev. A}, 99:022103, Feb 2019.

\bibitem{Karlovets2019Dynamical}
Dmitry Karlovets.
\newblock Dynamical enhancement of nonparaxial effects in the electromagnetic
  field of a vortex electron.
\newblock {\em Phys. Rev. A}, 99:043824, Apr 2019.

\bibitem{Landau1930}
D.~Landau.
\newblock Diamagnetismus der metalle.
\newblock {\em Z. Phys.}, 64:629--637, Sep 1930.

\bibitem{Ciftja2020}
O.~Ciftja.
\newblock Detailed solution of the problem of landau states in a symmetric
  gauge.
\newblock {\em European Journal of Physics}, 41(3):035404, Apr 2020.

\bibitem{MalkinManko}
I.~Malkin and V.~Manko.
\newblock {\em Dynamic symmetry and coherent states of quantum systems}.
\newblock Izdatel'stvo Nauka, Moscow, 1979.

\bibitem{BagrovGitman}
D.~Gitman V.~Bagrov.
\newblock {\em The Dirac equation and its solutions}.
\newblock Berlin [a. o.] : de Gruyter, 2014.

\bibitem{Filina2023}
N.~V. Filina and S.~S. Baturin.
\newblock {Unitary equivalence of twisted quantum states}.
\newblock {\em Phys. Rev. A}, 108(1):012219, 2023.

\bibitem{Siegman}
A.E. Siegman.
\newblock {\em Lasers}.
\newblock Mill Valley, 1986.

\bibitem{Feng2001}
S.~Feng and H.G. Winful.
\newblock Physical origin of the gouy phase shift.
\newblock {\em Optics Letters}, 26:485--487, Apr 2001.

\bibitem{Grillo2015}
V.~Grillo, G.~Gazzadi, E.~Mafakheri, S.~Frabboni, E.~Karimi, and R.~Boyd.
\newblock Holographic generation of highly twisted electron beams.
\newblock {\em Phys. Rev. Lett.}, 114:034801, Jan 2015.

\bibitem{Mafakheri2017}
E.~Mafakheri, A.~Tavabi, P.-H. Lu, R.~Balboni, F.~Venturi, C.~Menozzi, G.C.
  Gazzadi, S.~Frabboni, A.~Sit, R.~E. Dunin-Borkowski, E.~Karimi, and
  V.~Grillo.
\newblock Realization of electron vortices with large orbital angular momentum
  using miniature holograms fabricated by electron beam lithography.
\newblock {\em Appl. Phys. Lett.}, 110:093113, Mar 2017.

\bibitem{McMorran2017}
B.~McMorran, A.~Agrawal, P.~Ercius, V.~Grillo, A.~Herzing, T.~Harvey, M.~Linck,
  and J.~Pierce.
\newblock Origins and demonstrations of electrons with orbital angular
  momentum.
\newblock {\em Phil. Trans. R. Soc.}, 375:20150434, Feb 2017.

\bibitem{Gradshtein}
I.S. Gradshtein and I.M. Ryzhik.
\newblock {\em Table of Integrals, Series, and Products}.
\newblock FizMatGiz, Moscow, 4 edition, 1964.

\bibitem{Groening2021}
L.~Groening, C.~Xiao, and M.~Chung.
\newblock Particle beam eigenemittances, phase integral, vorticity, and
  rotations.
\newblock {\em Phys. Rev. Accel. Beams}, 24:054201, May 2021.

\bibitem{Reiser}
M.~Reiser.
\newblock {\em Theory and Design of Charged Particle Beams}.
\newblock Wiley, New York, 2008.

\bibitem{PhysRevSTAB.16.011302}
M.~Migliorati, A.~Bacci, C.~Benedetti, E.~Chiadroni, M.~Ferrario, A.~Mostacci,
  L.~Palumbo, A.~R. Rossi, L.~Serafini, and P.~Antici.
\newblock Intrinsic normalized emittance growth in laser-driven electron
  accelerators.
\newblock {\em Phys. Rev. ST Accel. Beams}, 16:011302, Jan 2013.

\bibitem{Noble1989}
R.J. Noble.
\newblock Beam mismatch and emittance oscillations in magnetic transport lines.
\newblock {\em Proceedings of the 1989 IEEE Particle Accelerator Conference},
  pages 1067 -- 1069, 1989.

\bibitem{Schrodinger1930}
E.~Schr{\"o}dinger.
\newblock {\em Zum Heisenbergschen Unsch{\"a}rfeprinzip}.
\newblock Abhandlungen der Preussischen Akademie der Wissenschaften,
  Physikalisch-Mathematische Klasse. Akademie der Wissenschaften, 1930.

\bibitem{Bagrov2014}
V~G Bagrov, D~M Gitman, and A~S Pereira.
\newblock Coherent and semiclassical states of a free particle.
\newblock {\em Physics-Uspekhi}, 57(9):891, sep 2014.

\bibitem{Hall}
Brian~C Hall.
\newblock {\em Quantum theory for mathematicians}.
\newblock Springer, 2013.

\end{thebibliography}

\end{document}